\documentclass[12pt]{article}
\usepackage[utf8]{inputenc}
\usepackage{geometry}
\usepackage{graphicx}
\usepackage{array}
\usepackage{slashed}
\usepackage{amsmath}
\usepackage{amsfonts}
\usepackage{amssymb}
\usepackage{physics}
\usepackage{mathtools}
\usepackage{placeins}
\usepackage[usenames, dvipsnames]{color}
\usepackage{caption}
\usepackage{subcaption}
\usepackage{dsfont}
\usepackage[cm]{fullpage}
\usepackage{cite}
\usepackage{epstopdf}
\usepackage{comment}
\usepackage{hyperref}
\usepackage{enumitem}
\usepackage{cancel}
\usepackage{mathrsfs}

\usepackage{cleveref}
\crefname{section}{§}{§§}
\Crefname{section}{§}{§§}
\usepackage{booktabs} 

\numberwithin{equation}{section}

\def\p{\partial}

\def\0{{(0)}}
\def\1{{(1)}}
\def\2{{(2)}}

\def\n{\nabla}

\def\<{\langle }
\def\>{\rangle }

\newcommand{\bea}{\begin{eqnarray}}

\newcommand{\eea}{\end{eqnarray}}

\newcommand{\be}{\begin{equation}}
\newcommand{\ee}{\end{equation}}
\newcommand{\ba}{\begin{align}}
\newcommand{\ea}{\end{align}}

   \makeatletter
  \let\over=\@@over \let\overwithdelims=\@@overwithdelims
  \let\atop=\@@atop \let\atopwithdelims=\@@atopwithdelims
  \let\above=\@@above \let\abovewithdelims=\@@abovewithdelims
\renewcommand\section{\@startsection {section}{1}{\z@}%
                                   {-3.5ex \@plus -1ex \@minus -.2ex}
                                   {2.3ex \@plus.2ex}%
                                   {\normalfont\large\bfseries}}

\renewcommand\subsection{\@startsection{subsection}{2}{\z@}%
                                     {-3.25ex\@plus -1ex \@minus -.2ex}%
                                     {1.5ex \@plus .2ex}%
                                     {\normalfont\bfseries}}

\newcommand{\beq}{\begin{equation}}
\newcommand{\eeq}{\end{equation}}
\newcommand{\beqa}{\begin{eqnarray}}
\newcommand{\eeqa}{\end{eqnarray}}
\newcommand{\beqar}{\begin{eqnarray*}}

\def\[{\[}
\def\]{\]}

\newcommand{\bd}[1]{\begin{fmffile}{#1}\begin{fmfgraph*}}
\newcommand{\ed}{\end{fmfgraph*}\end{fmffile}}

\begin{document}

\begin{titlepage}


\unitlength = 1mm~\\
\vskip 1cm
\begin{center}

{\LARGE{\textsc{Super No-Scale Models  \\[0.3cm] with Pati-Salam Gauge Group}}}

\vspace{0.8cm}
Ioannis Florakis\,{}\footnote{{\tt iflorakis@uoi.gr}}, John Rizos\,{}\footnote{\tt irizos@uoi.gr} and Konstantinos Violaris-Gountonis\,{}\footnote{\tt k.violaris@uoi.gr}

\vspace{1cm}

{\it  Department of Physics, University of Ioannina\\ GR45110 Ioannina, Greece 

}

\vspace{0.8cm}

\begin{abstract}
We construct and classify heterotic models where $\mathcal N=1$ supersymmetry is spontaneously broken \`a la Scherk-Schwarz. The theories contain chiral matter transforming under an observable Pati-Salam gauge symmetry and are classified according to a minimal set of phenomenological requirements, including the shape of their one-loop effective potential. The latter is essentially controlled by two parameters that are easily computable in terms of the defining string data. By imposing Bose-Fermi degeneracy at the massless level, known as the super no-scale condition, it is possible to suppress the value of the effective potential to exponentially small values, provided the compactification scale is sufficiently lower than the string scale. Furthermore, we compare the space of $\sim 10^{10}$ Pati-Salam models with their parent theories based on $SO(10)$ gauge symmetry and confirm the universality structure of their effective one-loop potentials.

\end{abstract}

\setcounter{footnote}{0}

\vspace{1.0cm}

\end{center}

\end{titlepage}

\pagestyle{empty}
\pagestyle{plain}

\def\vx{{\vec x}}
\def\p{\partial}
\def\po{$\cal P_O$}

\pagenumbering{arabic}

\tableofcontents

\section{Introduction}
Target space supersymmetry is one of the main properties commonly arising in constructions of consistent vacua within the framework of perturbative string theory. Its presence is intimately linked to the stability of the perturbative expansion at tree level as well as at higher genera. It is a well-known fact that the quantization of strings in a Minkowski background naturally leads to a vanishing value for the cosmological constant. Moreover, depending on the number of supercharges present in the theory, non-renormalisation theorems may drastically restrict quantum corrections in various terms in the string effective action.

Unfortunately, in non-supersymmetric string constructions most of these appealing features are no longer present. The effective action, including the scalar potential, is no longer super-protected and typically receives quantum corrections to all orders in perturbation theory. Generically, the cosmological constant now acquires a large non-vanishing value and reflects the presence of a dilaton tadpole already at one loop, which back-reacts against the tree-level background of the theory \cite{Fischler:1986tb,Fischler:1986ci}.  In addition, the spectrum is generically plagued by tachyonic excitations which induce IR instabilities and essentially invalidate a perturbative analysis \cite{Atick:1988si}.

If supersymmetry is realized in nature, it must be broken spontaneously and the determination of the precise mechanism as well as the scale of the breaking, within a fully-fledged stringy setup, remain important open problems. Non-supersymmetric string constructions have received renewed interest in the recent literature both in the open \cite{Abel:2018zyt,Angelantonj:1999gm,Angelantonj:2019gvi,Parameswaran:2020ukp,Coudarchet:2021qwc,Abel:2020ldo} (see also \cite{Mourad:2017rrl} for a recent review) as well as the closed string sector \cite{Angelantonj:2006ut,Gato-Rivera:2007ifz,Faraggi:2009xy,Angelantonj:2010ic,Blaszczyk:2014qoa,Lukas:2015kca,Ashfaque:2015vta,Blaszczyk:2015zta,Aaronson:2016kjm,Abel:2017vos,Kounnas:2017mad,Florakis:2018xma,Rizos:2018asy,Florakis:2020rph,Faraggi:2020wld,Itoyama:2020ifw,Perez-Martinez:2021zjj,Itoyama:2021fwc}. In the case of closed string theories, a way to break supersymmetry spontaneously while retaining the perturbative world-sheet description is the stringy version \cite{Rohm:1983aq,Kounnas:1988ye,Ferrara:1988jx,Kounnas:1989dk} of the Scherk--Schwarz mechanism \cite{Scherk:1978ta,Scherk:1979zr}. In its simplest form, it corresponds to introducing non-trivial monodromies to fields or vertex operators of the theory along compact cycles of the internal space, $\Phi(x,y+2\pi R) =e^{iQ}\Phi(x,y)$, associated to a symmetry generator $Q$ that is identified with the spacetime fermion number. This results in a shift of the Kaluza-Klein tower of states charged under $Q$ and introduces a mass gap controlled by the size $R$ of the internal cycle. States within the same supermultiplet are hence assigned different boundary conditions and the resulting theory corresponds to a spontaneous breaking of supersymmetry at a scale $m_{3/2}\sim 1/R$.

From the effective supergravity perspective, the Scherk-Schwarz mechanism corresponds to a flat gauging generating a tree-level scalar potential of the no-scale type. Its striking feature is that, once minimised with respect to the charged moduli, the potential vanishes, leaving the gravitino mass scale $m_{3/2}$ essentially undetermined at tree level. Strings living at this minimum can then be exactly quantized as freely-acting orbifolds, and the no-scale moduli controlling the supersymmetry breaking scale can be seen to correspond to exact marginal deformations of the world-sheet CFT.

The fate of the no-scale moduli is then determined by radiative corrections to the effective potential. Following the discussion of \cite{Florakis:2016ani}, if the one-loop potential contains a region of positive values $V_{1}(R_0) >0$ at radii sufficiently close to the string scale $R=R_0\sim 1$ (in string units $\alpha'=1$), while preserving its monotony properties for $R>R_0$, then the theory will be dynamically driven to large values of the no-scale modulus $R$. Since, in the present setup the supersymmetry breaking scale is of the order of the Kaluza-Klein scale $m_{3/2}\sim M_{KK}\sim 1/R$, one is naturally led to a large volume regime \cite{Antoniadis:1990ew}, while ensuring the separation of scales $m_{3/2} \ll M_{s}$.

A number of appealing features can be associated to the above scenario including, in particular, the protection of the theory against the excitation of tachyonic modes, as well as a considerable improvement of the stability properties of the perturbative expansion by means of the suppression of the dilaton tadpole. As discussed in \cite{Harvey:1998rc,Angelantonj:1999gm,Shiu:1998he, Abel:2015oxa, Kounnas:2016gmz, Florakis:2016ani}, the latter may be suppressed further by imposing a bose-fermi degeneracy to the massless degrees of freedom of the theory. The resulting theories, termed `super no-scale' models in \cite{Kounnas:2016gmz}, are characterised by an exponentially suppressed value of the cosmological constant at large radii, which could in principle be tuned to match observation after non-perturbative effects are taken into account in order to stabilise the remaining no-scale moduli (for a recent discussion, see \cite{Abel:2016hgy}). Note that obtaining an exactly vanishing value for the cosmological constant in non-supersymmetric heterotic orbifolds turns out to be a rather difficult problem \cite{GrootNibbelink:2017luf}.

In recent work \cite{Florakis:2016ani}, we set the ground for a systematic construction of heterotic theories in four dimensions, with chiral matter in representations of an SO(10) gauge group factor, where $\mathcal N=1$ supersymmetry is spontaneously broken via freely-acting orbifolds. There, the morphology of the one-loop corrected scalar potential was examined in various models and it was understood that, although the large volume asymptotics is directly controlled by the degeneracies of the light, physical states of the theory\footnote{For super no-scale models, this includes also the degeneracies of the first massive excitations.}, regions in moduli space of the order of the string scale are dominated by the contributions of massive and even non level-matched string states. 

In the large volume regime the physics becomes effectively higher dimensional and an additional problem emerges. Corrections to gauge couplings from the infinite tower of massive string modes exhibit linear growth with the compactification volume. Depending on the sign of the corresponding beta function, the relevant gauge group factor then either decouples or is driven quickly to the non-perturbative regime, and predictability is essentially lost. This is the so-called decompactification problem \cite{Antoniadis:1990ew,Kiritsis:1996xd,Faraggi:2014eoa}, which is not particular to a given construction but, rather, a generic feature of the separation of the Kaluza-Klein and string scales.

The behaviour of running gauge couplings within our framework was investigated in \cite{Florakis:2017ani}, where the origin of the decompactification problem was traced down to properties of a parent $\mathcal N=1$ theory in six dimensions. One of the upshots of our analysis was the reinterpretation of the decompactification problem as a vacuum selection principle, giving rise to precise group-theoretic conditions to be imposed upon the parent 6d theory, which remove the problematic linear growth and retain only a moderate logarithmic dependence on the volume of the compactified space, while allowing the presence of chiral matter. Moreover, as was recently discussed in \cite{Angelantonj:2018gpm}, the absence of the decompactification problem is intricately related to the possibility of unifying gauge couplings at a scale compatible with bottom-up GUT scenarios. Therefore, it becomes possible and important to study the running of gauge couplings at scales much lower than the string scale and investigate questions of unification within this framework.

Albeit useful for a detailed study of the shape of the effective one-loop potentials as well as for identifying a solution to the decompactification problem while preserving chirality, the constructions considered in \cite{Florakis:2016ani,Florakis:2017ani} are not suitable for the investigation of further phenomenological properties of interest. For instance, although containing chiral matter organised in representations of an `observable' SO(10) GUT gauge group, the Higgs sector responsible for its eventual breaking to the Standard Model is absent in those theories. Moreover, the presence of a single non-abelian SO(10) factor precludes a study of the running of different observable gauge couplings to energy scales much lower than $M_s$ and investigating related questions of unification. In addition, the results obtained in \cite{Florakis:2016ani,Florakis:2017ani} were based on a random scan of $10^{11}$ models subject to a set of preliminary conditions, but no exhaustive classification of the theories or their effective potentials was attempted.

In the present work we aim to address these questions, by systematically constructing heterotic four-dimensional super no-scale models with spontaneously broken $\mathcal N=1\to 0$ supersymmetry, and with chiral matter charged under an observable Pati-Salam gauge group $SU(4)_C\times SU(2)_L\times SU(2)_R$. The models are first constructed at a special point in the perturbative moduli space, in which the $\hat c=6$ SCFT of internal coordinates is conveniently replaced by an equivalent CFT, entirely realized in terms of free world-sheet fermions \cite{Antoniadis:1986rn,Antoniadis:1987wp}. In fact, the Pati-Salam gauge symmetry \cite{Pati:1974yy} is the simplest choice containing three non-abelian gauge group factors that still admits a string-theoretic realisation \cite{Antoniadis:1988cm,Antoniadis:1990hb,Leontaris:1999ce} in terms of free-fermions with purely real boundary conditions. This allows for a powerful and exhaustive computer aided scan of the corresponding theories, which are subsequently moved away from the fermionic point by means of marginal deformations of the current-current type, $J_L(z)\times \tilde J_R(\bar z)$.

The latter deformed theories are then explicitly matched with corresponding compactifications on toroidal $T^6/(\mathbb Z_2)^2 \times (\mathbb Z_2)^n$ orbifolds, where the first two $\mathbb Z_2$ factors generate the singular limit of $\mathcal N=1$ Calabi-Yau compactifications with standard embedding and $SO(10)$ observable gauge group, whereas the remaining $\mathbb Z_2$ factors are responsible for the spontaneous breaking of supersymmetry, as well as the reduction of the gauge group down to the Pati-Salam group. It is then possible to study explicit properties of these orbifold theories, such as the presence of chiral matter, 
the presence of symmetry breaking Higgs scalars and, importantly, to exhaustively classify the structure of the one-loop effective potential at the generic point in the Scherk-Schwarz moduli space.

A priori, our construction gives rise to $\sim 7\times 10^{13}$ models, though this number can be considerably reduced by imposing a minimal number of preliminary constraints. Specifically, we require that, despite the Scherk-Schwarz breaking of spacetime supersymmetry, no tachyonic excitations be encountered at the fermionic point. We further require the super no-scale property, necessary to suppress the cosmological constant at large volume, as well as the presence of complete generations of chiral matter under the observable $SU(4)_C\times SU(2)_L\times SU(2)_R$ gauge group. Additional conditions arise by demanding that  Standard Model Higgs doublets survive in the massless spectrum of the theory. Similarly, the presence of at least one additional Higgs field is necessary for the eventual breaking of the Pati-Salam group down to the Standard Model.

It is important to stress that the above conditions are non-trivial and it is a priori far from obvious whether models satisfying all the above constraints actually exist. Moreover, an isolated analysis conducted purely at the fermionic-point is unable to distinguish between spontaneous and explicit breaking of supersymmetry. Indeed, additional constraints are required \cite{Florakis:2016ani} in order for the breaking to be consistent with the Scherk-Schwarz realization in terms of freely-acting orbifolds at generic points of the perturbative moduli space.

The purpose of this work is to show that the space of viable models passing the aforementioned tests is not empty, and then proceed to exhaustively classify their massless spectra as well as the structure of their effective potentials. The main observation arising from our analysis is that the super no-scale structure is not an accidental feature of special toy-models but, rather, a property which can generically persist in more interesting classes of semi-realistic constructions. Furthermore, a direct comparison between the super no-scale models of \cite{Florakis:2016ani} based on SO(10), and those enjoying the Pati-Salam gauge symmetry constructed here, reveals that the salient characteristics of the morphology of the one-loop potentials are largely unchanged. Similarly, the number of chiral matter generations present in the theory does not appear to significantly affect or restrict the general structure of the potentials.
		
This paper is organised as follows. In Section $2$, we set the ground for the construction of heterotic Pati-Salam models within the free fermionic framework, with special emphasis on the self-consistency conditions for the absence of tachyonic excitations. We discuss the massless spectrum of the theory and outline the core phenomenological features that will play a role in the subsequent classification of the theories, with emphasis on the properties of chirality, as well as the presence of heavy and light Higgs bosons. Section $3$ reintroduces the dependence on the geometric moduli, by rewriting the models as toroidal orbifold compactifications. We then proceed to discuss the structure of the one loop effective potential and explain how the qualitative features of its shape are effectively controlled by a minimal number of parameters. Section $4$ contains the main results of our exhaustive computer-aided scan of the space of models. We outline the main phenomenological criteria of interest, and proceed to classify the theories into well-defined categories, while taking also their one-loop effective potentials into account. In Section $5$ we give a detailed analysis of a number of specific models and discuss some of their features. 	We end in Section $6$ with our conclusions.



\section{Non-supersymmetric Pati--Salam models\label{section2}}

In this section, we define the class of non-supersymmetric Pati--Salam heterotic string models under consideration and investigate their potential tachyonic modes as well as their massless spectra. We also formulate a set of necessary requirements that guarantee model consistency and compatibility with low energy phenomenology, including the absence of tachyons, the existence of chiral fermions and the presence of scalar Higgs fields associated with Pati--Salam (PS) and SM gauge symmetry breaking. The free fermionic formulation 
of heterotic strings \cite{Antoniadis:1986rn,Antoniadis:1987wp,Kawai:1986va}  provides a convenient framework for our analysis as it allows the formulation of these requirements in terms of generalised
GSO projections. This in turn permits a systematic exploration and classification of large sets of four-dimensional string models utilising the methodology developed in \cite{Gregori:1999ny,Faraggi:2004rq,Faraggi:2006bc}.

In the free fermionic formulation of the heterotic string in 4d all world-sheet bosonic coordinates are fermionised except for those associated with four-dimensional space-time. In the standard notation, the world-sheet fermionic coordinates comprise 20 real left-moving fermions $\big\{\psi^\mu$, $\chi^{12},\chi^{34},\chi^{56}$, $y^{12},y^{34}, y^{56}, \omega^{12},\omega^{34},\omega^{56}\big\}$ along with 12 real and 16 complex right-moving fermions $\big\{\bar{y}^{12},\bar{y}^{34}, \bar{y}^{56},\bar{\omega}^{12},\bar{\omega}^{34},\bar{\omega}^{56},$
$\bar{\psi}^{1,\dots,5},\bar{\eta}^{1,2,3},\bar{\phi}^{1,\dots,8}\big\}$.
Here, $\psi^\mu$ describe the space-time fermions in the light-cone gauge,
$\chi^k, k=1,\dots,6$ are the remaining RNS fermions, $y^I,\omega^I,\bar{y}^I,\bar{\omega}^I$, $I=1,\dots6$, parametrise the  internal  left-right moving coordinates upon fermionisation, and 
$\bar{\psi}^a,\bar{\eta}^{b},\bar{\phi}^{c}$, $a=1,\dots,5$, $b=1,2,3$, $c=1,\dots,8$ are the 16 right-moving degrees of freedom in the fermionic realisation of the heterotic string.
In this framework a model is defined by a set of basis vectors which encode the parallel transport properties of the 2d fermionic fields along the non-contractible loops of the world-sheet torus, and a set of phases
 associated with generalised GSO projections (GGSO).

The class of models discussed here is defined by the following set of ten basis vectors of boundary conditions $\{\beta_1,\beta_2,\dots,\beta_{10}\}$:
\begin{align}
\begin{split}
\beta_1=\mathds{1}&=\{\psi^\mu,\
\chi^{1,\dots,6},y^{1,\dots,6},\omega^{1,\dots,6}|\bar{y}^{1,\dots,6},
\bar{\omega}^{1,\dots,6},\bar{\eta}^{1,2,3},
\bar{\psi}^{1,\dots,5},\bar{\phi}^{1,\dots,8}\}\\
\beta_2=S&=\{\psi^\mu,\chi^{1,\dots,6}\}\\
\beta_3=T_1&=\{y^{12},\omega^{12}|\bar{y}^{12},\bar{\omega}^{12}\}\\
\beta_4=T_2&=\{y^{34},\omega^{34}|\bar{y}^{34},\bar{\omega}^{34}\}\\
\beta_5=T_3&=\{y^{56},\omega^{56}|\bar{y}^{56},\bar{\omega}^{56}\}\\
\beta_6=b_1&=\{\chi^{34},\chi^{56},y^{34},y^{56}|\bar{y}^{34},\bar{y}^{56},
\bar{\psi}^{1,\dots,5},\bar{\eta}^1\}\\
\beta_7=b_2&=\{\chi^{12},\chi^{56},y^{12},y^{56}|\bar{y}^{12},\bar{y}^{56},
\bar{\psi}^{1,\dots,5},\bar{\eta}^2\} \\
\beta_8=z_1&=\{\bar{\phi}^{1,\dots,4}\} \\
\beta_9=z_2&=\{\bar{\phi}^{5,\dots,8}\} \\
\beta_{10}=\alpha&=\{\bar{\psi}^{4,5},\bar{\phi}^{1,2}\} ,
\end{split}\label{basis}
\end{align}	
where included fermions are periodic and the rest anti-periodic, together with a set of phases $c[^{\beta_i}_{\beta_j}]=\pm1$, $i,j=1,2,\dots,10$ subject to
restrictions  dictated by modular invariance. Vectors $\mathds{1}, S, T_1, T_2, T_3$ define an $N=4$ supersymmetric model exhibiting ${SO(4)}^3{\times}SO(32)$
gauge symmetry. Vectors $b_1, b_2$ correspond to the $Z_2\times{Z_2}$ orbifold twists with standard embedding which reduce supersymmetry to $N=1$ and the gauge symmetry to 
${U(1)}^6\times{SO(10)}\times{U(1)}^2\times{SO(18)}$, where $SO(10)$ plays the role of the observable gauge group. In this language $T_1, T_2, T_3$ are associated with orbifold shifts along the three internal 2-tori. The addition
of $z_1, z_2$  reduces the hidden $SO(18)$ gauge symmetry to
${U(1)}\times{SO(8)}^2$, while $\alpha$ breaks the observable $SO(10)$ group to
the Pati--Salam gauge symmetry $SO(6)\times SO(4)\sim SU(4)\times{SU(2)}_L\times{SU(2)}_R $ and further reduces the hidden gauge group ${SO(8)}^2{\to}{SO(4)}^2\times{SO(8)}\sim SU(2)^4\times SO(8)$. Altogether, the models
possess 
\begin{equation}\label{Gauge Symmetry}
\begin{aligned}
G&= \left\{SU(4)\times SU(2)_L\times SU(2)_R\right\}_{\rm observable}\times U(1)^3\times SU(2)^4\times SO(8)
\end{aligned}
\end{equation}
gauge symmetry apart from special choices of GGSO phases which allow for further gauge group enhancements, and we have suppressed a standard ${U(1)}^6$ factor originating from the compactification. 

As far as the GGSO phases $c_{ij}=c{\beta_i\atopwithdelims[]\beta_j}$ are concerned, we can use modular invariance to fix all $c_{ij}$ for $i>j=1,\dots,10$ and $c_{ii},i=2,\dots,10$ in terms of $c_{11}, c_{ij}, i<j=1,\dots,10$. Moreover,
in the free fermionic framework space-time SUSY breaking can be realised by projecting out the gravitino state in the sector $S$. For this purpose, we employ, without loss of generality, a GGSO projection associated with the first 2-torus
\begin{align}
c_{23}=c{S\atopwithdelims[]T_1}=+1\,.
\end{align}
As it will become clear in Section \ref{section3}  where we discuss the implementation of the Scherk--Schwarz mechanism  this 
is compatible with a mass for the gravitino inversely
proportional to the compactification radius. 
In total, we are left with $10(10-1)/2=45$ independent phases that give rise to $2^{45}\sim 3.5\times 10^{13}$ a priori distinct non-supersymmetric Pati--Salam models. However, a 
number of these models do not correspond to consistent string vacua  due to the presence of tachyonic instabilities already at tree-level.

The spectrum of a  string model generated by the basis vectors \eqref{basis} is organised in $2^{10}$ sectors defined as
\begin{align}
\sum_{i=1}^{10} m_i \beta_i\quad,\quad m_1,\dots,m_{10} = \left\{0,1\right\}\,.
\end{align}
In this framework, level-matched tachyons with $M^2=-1/2$ and/or $M^2=-1/4$
can be encountered in $19$ sectors.
Whether such physical tachyons appear in a given sector is determined by a projection operator expressible explicitly in terms of the GGSO coefficients. 
Introducing the generalised projectors
\begin{align}
\mathds{P}_a^\pm= \prod_{\xi\in\Xi^\pm(a)}\frac{1}{2}\left(1\pm c{a\atopwithdelims[]{\xi}}^\ast\right)\ ,\ 
\label{projj}
\end{align}
the composite vector
$
x=\mathds{1}+S+T_1+T_2+T_3+z_1+z_2=\left\{\bar{\psi}^{1,\dots,5},\bar{\eta}^1,\bar{\eta}^2,\bar{\eta}^3\right\}
$ and $b_3=b_1+b_2+x$, 
the projection operators signalling the existence of $M^2=-1/2$ tachyons, generated in the $a \in \left\{z_1, z_2, \alpha, z_1+\alpha\right\}$ sectors are $\mathds{P}_a^+$, where
\begin{align}
\Xi^+(z_1)&=\left\{S,T_1, T_2, T_3, b_1, b_2, z_2\right\}\nonumber\\
\Xi^+({z_2})&=\left\{S,T_1, T_2, T_3, b_1, b_2, z_1, \alpha\right\}\\
\Xi^+(\alpha)&=\left\{S,T_1, T_2, T_3, b_1+b_2, b_1+z_1+x, z_2\right\}\nonumber\\ 
\Xi^+({z_1+\alpha})&=\left\{S,T_1, T_2, T_3, b_1+b_2,  b_1+\alpha, z_2\right\}\,.\nonumber
\end{align}
Tachyons with mass $M^2=-1/4$ can be generated  for 
$a\in\left\{T_i +  z_1, T_i +  z_2\right.$, $\left. T_i + p z_1+\alpha\right\}$, with $i=1,2,3$ and $p=0,1$. The associated projector operators are $\mathds{P}_a^+$ with
\begin{align}
\Xi^+(T_i+z_1) & = \left\{S,T_j, T_k, b_i, z_2\right\}\nonumber\\
\Xi^+(T_i+z_2) & = \left\{S,T_j, T_k, b_i, z_1, \alpha\right\}\\
\Xi^+(T_i+\alpha) & = \left\{S,T_j, T_k, b_i+x, z_1+\alpha+x, z_2\right\}\nonumber\\
\Xi^+(T_i+z_1+\alpha) & = \left\{S,T_j, T_k, b_i+x, b_i+\alpha, z_2\right\}\,,
\nonumber
\end{align}
 where $i=1,2,3, i\ne j\ne k$.
Furthermore, tachyons with mass $M^2=-1/4$ arise from the sectors $T_1, T_2, T_3$ upon the 
action of right-moving fermion oscillators. 
In this case, the projection operators regarding physical tachyons are more complicated. In order to describe them we introduce a 
 a general projector depending on a set of vectors $\Xi(a)$, each of which may or may not include a specific fermion $\varphi$, on states resulting from sector $a$  
\begin{align}
\mathds{P}_a^\varphi= \prod_{\xi\in\Xi(a)}\frac{1}{2}\left(1+\delta^\varphi_{\xi}c{a\atopwithdelims[]{\xi}}^\ast\right)\,,
\end{align}
with 
\begin{align}
\delta^\varphi_a = 
\begin{cases}
-1\ ,\ \varphi\in a\\
+1\ ,\ \varphi\notin a
\end{cases}
\ .
\end{align}
Then, 
the tachyonic states coming from the sectors $T_i, i=1,2,3$ can be recast in the form
\begin{equation}
\left|\left(y^{2i-1,2i}\omega^{2i-1,2i}\right)\right>_L\times
\left(\bar{\phi}_{-\frac{1}{2}}
\atop
\bar{\phi}_{-\frac{1}{2}}^\ast
\right)
\left|\left(\bar{y}^{2i-1,2i}\bar{\omega}^{2i-1,2i}\right)\right>_R\ ,\ 
\bar{\phi}\in \mathds{1}_R-{T_{i}}_R\,,
\end{equation}
where ${\beta_i}_R$ denotes the subset of right-moving fermions of the basis vector $\beta_i$, and
the associated projectors are $\mathds{P}_{T_i}^{\bar\varphi}$ with
\begin{align}
\Xi(T_i) = \left\{S,T_j,T_k,b_i,z_1,z_2,\alpha\right\}\ , i\ne j \ne k\,.
\end{align}
In order for our models to be free of physical tachyons, we need to make sure that all tachyonic sector projection operators vanish
\begin{equation}
\label{tachyonprojectors}
\begin{split}
\mathds{P}_a^+ &= 0\ ,\ a \in \left\{z_1,z_2,\alpha,z_1+\alpha\right\}\cup
\left\{T_i +  z_1, T_i + z_2, T_i + p z_1+\alpha\right\}\ ,\\
&\qquad\qquad\qquad\qquad\qquad\qquad\qquad\qquad\qquad\  i=1,2,3,\ p=0,1\,,\\
\mathds{P}^{\bar{\phi}}_{T_j}&=0\ ,\ \bar{\phi}\in \mathds{1}_R-{T_{j}}_R\ ,\ j=1,2,3\,.
\end{split}
\end{equation}
Note that physical tachyons carrying oscillators of the right-moving internal coordinates $\left\{\bar{y}^{12}, \bar{y}^{34}, \bar{y}^{56}\right.$, $\left.\bar\omega^{12}, \bar\omega^{34}, \bar\omega^{56}\right\}$ disappear from the spectrum when moving away from the fermionic point. However, we choose to project them out in order to ensure consistency of the spectrum at the fermionic point which is the starting point of our analysis.

Let us focus on the massless spectrum of the class of string models under consideration. First note that the  Pati--Salam model we employ here
is a variant of the original Pati--Salam ${SU(4)}\times{SU(2)_L}\times{SU(2)_R}$ model \cite{Pati:1974yy} based on \cite{Antoniadis:1988cm}. The reason for this is that in the original model the breaking of the gauge symmetry to that of the SM is realised using Higgs scalars in the adjoint representation. These are not available in the context of $k=1$ Kac--Moody realisations of the gauge symmetry on the string world sheet as the ones considered here. Therefore, our PS models utilise the standard assignments of the Standard Model fermions with the addition of a right-handed neutrino
\begin{equation*}
\begin{aligned}
&F_L({\bf4},{\bf2},{\bf 1})= Q({\bf3},{\bf2},-1/6)+L({\bf1},{\bf2},1/2)\,,\\
&\overline{F}_R({\bf\bar 4},{\bf 1},{\bf 2})= u^c({\bf\bar 3},{\bf1},2/3)+d^c({\bf\bar 3},{\bf1},-1/3)+e^c({\bf 1}, {\bf 1},-1)+\nu^c({\bf 1}, {\bf 1},0)\,. \\
\end{aligned}
\label{apt}
\end{equation*}
These are accommodated in $SO(10)$ spinorials decomposed under $SU(4)\times SU(2)_L\times SU(2)_R$ as:
\begin{equation*}
\begin{aligned}
{\bf 16}=F_L({\bf 4}, {\bf 2}, {\bf 1})+\overline{F}_R({\bf \bar{4}}, {\bf 1}, {\bf 2})\,,
\end{aligned}
\end{equation*}
while $SO(10)$ anti-spinorials can accommodate anti-families 
\begin{equation*}
\begin{aligned}
{\bf\overline{16}}=\overline{F}_L({\bf \bar{4}},{\bf 2},{\bf 1})+F_R({\bf 4},{\bf 1},{\bf 2})\,. \\
\end{aligned}
\end{equation*}
Pati--Salam symmetry breaking scalars are assigned to 
\begin{align}
H\left({\bf4},{\bf1},{\bf 2}\right)+\text{h.c.}
\label{Hscalars}
\end{align}
 The neutral component of a complex scalar in the $H({\bf4}, {\bf1}, {\bf2})$+h.c. representation can acquire a VEV, leading to the spontaneous breaking of the Pati-Salam gauge symmetry. These PS symmetry breaking  bosons are necessary for our models to make contact with low energy physics.
 
In addition, the bi-doublet representation accommodates the SM breaking Higgs scalars 
\begin{align}
h\left({\bf1},{\bf2},{\bf 2}\right) = H_u\left({\bf1},{\bf2},+\frac{1}{2}\right)
+ H_d\left({\bf1},{\bf2},-\frac{1}{2}\right)\,.
\label{SMHiggs}
\end{align}
In this setting the SM fermion masses arise from a single type of coupling
\begin{align}
F_L \overline{F}_R h\,.
\end{align}
 
 The aforementioned states are the necessary ingredients for 
a minimal implementation of a non-supersymmetric low energy model based on the Pati-Salam gauge group. However, in the context of the string construction additional states can appear. These include, 
$SU(4)$ sextets 
$
\left({\bf6},{\bf1},{\bf 1}\right)\,
$
accommodating extra fermion triplets, 
fractionally charged fermionic exotic states in the representations
$
\left({\bf4},{\bf1},{\bf 1}\right)\ ,\ \left({\bf\bar4},{\bf1},{\bf 1}\right)
$
and
$
\left({\bf1},{\bf2},{\bf 1}\right)+\left({\bf1}, {{\bf 1},\bf2}\right)\,,
$ and/or bosonic/fermionic ``partners" of all the above-mentioned states. 

It turns out that a string model generated by the set of basis vectors \eqref{basis} comprises 229 sectors that could potentially give rise to massless modes. These include 122 bosonic and 107 fermionic sectors. Chiral  matter  is produced in the twisted sectors. More specifically, there are $12$ sectors,
 ${\cal S}^i_{pq}=S+b_i+p T_j+q T_k$,	where $p,q=0,1$,  $(i,j,k)=\{(1,2,3),(2,1,3),(3,1,2)\}$, which can produce 
 chiral and/or anti-chiral matter $F_L,\overline{F}_R$, $\overline{F}_L,{F}_R\,$.
  Each family or anti-family, in the context of these models, is obtained by combining the surviving states from two different sectors as the $\alpha$ related projections truncate the $SO(10)$ spinorials emerging from  ${\cal S}^i_{pq}$. The projections that determine whether chiral states survive the GGSO are
 \begin{equation}
 \label{spinorial fermion projectors}
{\mathds{P}}_{{\cal S}^i_{pq}}^-=1\ ,\ \Xi^-\left({\cal S}^i_{pq}\right)=\left\{T_i,z_1,z_2\right\}\,.
 \end{equation}
Using this we can write down expressions for the multiplicities $n_{L}, \bar{n}_{R}, \bar{n}_{L}, n_{R}$ of  $F_L,\overline{F}_R$, $\overline{F}_L,{F}_R\,$ respectively, as
 \begin{equation}
 \begin{aligned}
 &n_{L}=4 \sum_{i=1}^3\sum_{p,q=0}^1 \mathds{P}_{{\cal S}^i_{pq}}^-\frac{1}{2}\left(1+X_{{\cal S}^i_{pq}}^{SU(4)}\right)\frac{1}{2}\left(1+X_{{\cal S}^i_{pq}}^{SO(4)}\right)\,, \\
 &\bar{n}_{R}=4 \sum_{i=1}^3\sum_{p,q=0}^1 \mathds{P}_{{\cal S}^i_{pq}}^-\frac{1}{2}\left(1-X_{{\cal S}^i_{pq}}^{SU(4)}\right)\frac{1}{2}\left(1-X_{{\cal S}^i_{pq}}^{SO(4)}\right)\,,\\
 &\bar{n}_{L}=4 \sum_{i=1}^3\sum_{p,q=0}^1 \mathds{P}_{{\cal S}^i_{pq}}^-\frac{1}{2}\left(1-X_{{\cal S}^i_{pq}}^{SU(4)}\right)\frac{1}{2}\left(1+X_{{\cal S}^i_{pq}}^{SO(4)}\right)\,, \\
 &n_{R}=4\sum_{i=1}^3\sum_{p,q=0}^1 \mathds{P}_{{\cal S}^i_{pq}}^-\frac{1}{2}\left(1+X_{{\cal S}^i_{pq}}^{SU(4)}\right)\frac{1}{2}\left(1-X_{{\cal S}^i_{pq}}^{SO(4)}\right)\,, \\
 \end{aligned}
 \label{efga}
 \end{equation}
where $X_{{\cal S}^i_{pq}}^{SU(4)}$, $X_{{\cal S}^i_{pq}}^{SO(4)}$ stand for the $SU(4)$, $SO(4)\sim{SU(2)}_L\times{SU(2)}_R$ projectors
that distinguish between ${\mathbf4}/{\bar{\mathbf4}}$ and ${\mathbf2}_L=\left({\mathbf2},{\mathbf1}\right)/{\mathbf2}_R=\left({\mathbf1},{\mathbf2}\right)$ respectively 
\begin{equation}\label{spinorial fermion representation operators}
\begin{aligned}
&X_{{\cal S}^i_{pq}}^{SU(4)}=-c\begin{bsmallmatrix}
{\cal S}^{i}_{pq} \\ {\cal S}^j_{0,1-q}+\alpha \end{bsmallmatrix}^*\,,\,j \ne i=1,2\,
 ,\ 
X_{{\cal S}^3_{pq}}^{SU(4)}=-c\begin{bsmallmatrix}
{\cal S}^{3}_{pq} \\ {\cal S}^1_{1-q,0}+\alpha \end{bsmallmatrix}^*\ ,\ 
X_{{\cal S}^i_{pq}}^{SO(4)}=-c\begin{bsmallmatrix}
{\cal S}^i_{pq} \\ \alpha \end{bsmallmatrix}^*\,,i=1,2,3\,.
\end{aligned}
\nonumber
\end{equation}
 The net number of fermion generations can then be expressed as
 \begin{equation}
 n_g=n_{L}-\bar{n}_{L}= \bar{n}_{R}-n_{R}\,.
 \label{netchirality}
 \end{equation}
Due to the asymmetric action of the $\alpha$ vector related GGSO projection, the condition $n_{L}-\bar{n}_{L}= \bar{n}_{R}-n_{R}$ is not automatically satisfied and must be imposed as a constraint.

Fermions in spinorial representations can also arise from the $S+x$ sector. This sector, however,
being untwisted does not contribute to chirality. Actually, we can fix the $SO(4)$ projection utilising the $\alpha$ vector, but there is no way to differentiate between ${\mathbf4}$ and ${\bar{\mathbf4}}$ $SU(4)$ representations. This means that if $S+x$ does generate physical states,
these always come in pairs $\left({\bf4}+\overline{\bf4}, {\bf2}, {\bf1}\right)$ or 
 $\left({\bf4}+\overline{\bf4}, {\bf1}, {\bf2}\right)$ that do not change the 
 net chirality \eqref{netchirality}.
 
There are $13$ sectors that potentially give rise to scalars accommodated in  $SO(10)$ spinorial representations introduced above, $\mathbf{ \varSigma}^i_{pq}=b_i+pT_j+qT_k$, where $p,q=0,1$, $(i,j,k)=\{(1,2,3),(2,1,3),(3,1,2)\}$ and $x$. These are, of course, the bosonic partners of the aforementioned fermionic sectors. The corresponding projections on the $\mathbf{ \varSigma}^i_{pq}$ sectors  determining whether states survive GGSO are
\begin{align}
\mathds{P}_{{\mathbf\varSigma}^i_{pq}}^+\mathds{P}_{{\mathbf\varSigma}^i_{pq}}^-=1\,, \Xi^+\left({\mathbf\varSigma}^i_{pq}\right)=\left\{T_i,z_1,z_2\right\}\ ,\ 
\Xi^-\left({\mathbf\varSigma}^i_{pq}\right)=\left\{\alpha\right\}\,.
\end{align}
In the $x$ sector, massless states are constructed by acting on the vacuum with a left-moving oscillator $\phi_{-\frac{1}{2}}$, $\phi\in$  $\left\{\chi^1,\dots\chi^6,y^1,\dots,y^6,\omega^1,\dots,\omega^6\right\}$.
However, it turns out that these states are not relevant to our analysis. As a matter of fact, states involving $y,\omega$-oscillators disappear when departing from the fermionic point,  while states containing $\chi$-oscillators are always accompanied by states carrying $\psi^\mu$-oscillators which we choose to project out in order to avoid observable group enhancements. As a result, the number of PS breaking Higgs multiplets $n_H$ can be expressed as
\begin{align}
n_H = \sum_{i=1}^3\sum_{p,q=0}^1 \mathds{P}_{{\mathbf\varSigma}^i_{pq}}^+ \mathds{P}_{{\mathbf\varSigma}^i_{pq}}^-
\,.
\label{HPS}
\end{align}

Let us now consider the SM breaking scalar Higgs doublets \eqref{SMHiggs} residing in vectorial $SO(10)$ representations. In the models under  consideration these states arise  from the sectors ${\bf\varUpsilon}^i_{pq}={\bf\varSigma}^i_{pq}+x=b_i+pT_j+qT_k+x, i\ne j \ne k$, and/or $T_a+T_b$, $a\ne{b}=\{1,2,3\}$, by acting on the vacuum with a right-moving fermionic oscillator $\overline{\psi}^k_{-\frac{1}{2}}, k=4,5$. However, the Higgs scalars  generated from the sectors $T_a+T_b$ go away when departing from the fermionic point and will not be further considered. As a result,  the total number of scalar PS bi-doublets can be recast in the form
\begin{align}
n_h = \sum_{i=1}^3\sum_{p,q=0}^1 \mathds{P}^{\overline{\psi}^{4,5}}_{{\bf\varUpsilon}^i_{pq}}\,,
\label{HSM}
\end{align}
where $\Xi\left({\bf\varUpsilon}^i_{pq}\right)=\left\{T_i,z_1,z_2,\alpha\right\}$.

In the context of the Pati--Salam GUT the embedding of the electric charge generator is given by
\begin{equation}
Q_{em}=\frac{1}{\sqrt{6}}T_{15}+\frac{1}{2}I_{3_L}+\frac{1}{2}I_{3_R},
\end{equation}
where $T_{15}$, $I_{3_L}$ and $I_{3_R}$ are diagonal generators of $SU(4)$, $SU(2)_L$ and $SU(2)_R$ respectively. This leads to standard electric charge assignments for all massless states discussed so far. However, the spectrum of the string models contain also exotic fractionally charged colourless states that arise from sectors containing the $SO(10)$ breaking basis vector $\alpha$. The presence of these states is a generic feature of string compactifications \cite{Wen:1985qj,Athanasiu:1988uj,Schellekens:1989qb,Chang:1996vw,Coriano:2001mg}.

 One  example of exotic states are massless fermions in the representations $({\bf 4},{\bf 1},{\bf 1})$ or $({\bf \bar{4}},{\bf 1},{\bf 1})$ with respect to $SU(4)\times SU(2)_L\times SU(2)_R$, transforming as doublets of the hidden $SU(2)_1\times SU(2)_2=SO(4)_1$ or $SU(2)_3\times SU(2)_4=SO(4)_2$ gauge group. These states  arise from the sectors:
\begin{equation*}
E^{i}_{pqr}=\mathcal{S}^i_{pq}+rz_1+\alpha=S+b_i+pT_j+qT_k+rz_1+\alpha,
\end{equation*}
where $i\neq j\neq k$ and $p,q,r=0,1$.
The projection operators determining their presence are:
\begin{equation}
\mathds{P}^-_{E^{i}_{pqr}}=1,\;\Xi^-(	E^{i}_{pqr})=\{T_i,(1-r)z_1+\alpha,z_2\}\,.
\end{equation}
The number of exotic (anti-)quadruplets arising in these sectors can then be expressed as:
		\begin{equation}
		\begin{aligned}
		&n_{\mathbf4}=4\sum_{i=1,2,3\atop p,q,r=0,1}\mathds{P}^-_{E^{i}_{pqr}}\frac{1}{2}\left(1+X^{SU(4)}_{E^{i}_{pqr}}\right) \\
		&n_{\overline{\mathbf4}}=4\sum_{i=1,2,3\atop p,q,r=0,1}\mathds{P}^-_{E^{i}_{pqr}}\frac{1}{2}\left(1-X^{SU(4)}_{E^{i}_{pqr}}\right)\,,
		\end{aligned}
		\label{n44b}
		\end{equation}
where
	\begin{equation}
	X^{SU(4)}_{E^{i}_{pqr}}=\begin{cases}
	-c\left[^{E^{i}_{pqr}}_{\mathcal{S}^{j}_{0,1-q}}\right],\;i\neq j=1,2 \\
	-c\left[^{E^{i}_{pqr}}_{\mathcal{S}^{1}_{1-q,0}}\right],\;i=3
	\end{cases}
	\end{equation}
	distinguish between $\bf 4$/$\bf{\bar{4}}$.
In addition to these, exotic fermion quadruplets can also arise in the sectors $S+\alpha+x$, and $S+z_1+\alpha+x$. These untwisted sectors, however, can only generate pairs of $(\bf{4}+\bf{\bar{4}},\bf {1},\bf{1})$ and will not be taken into account in \eqref{n44b} for reasons that will become clear later in this section.

In addition, exotic fermions transforming as bi-doublets of $SU(2)_{L/R}\times SU(2)_{1/2/3/4}$ arise in the sectors:
\begin{equation*}
\mathcal{E}^{i}_{pqr}=\mathcal{S}^i_{pq}+x+rz_1+\alpha=S+b_i+pT_j+qT_k+x+rz_1+\alpha\;.
\end{equation*}
The corresponding projection operators are:
\begin{equation}
\mathds{P}^-_{\mathcal{E}^{i}_{pqr}}=1,\;\Xi^-(	\mathcal{E}^{i}_{pqr})=\{T_i,z_2,(1-p)T_j+(1-q)T_k+b_i+(1-r)z_1+\alpha\}\;.
\end{equation}
Fermion bi-doublets can also arise from the Ramond vacuum of the sectors $a=\{S+z_2+\alpha,S+z_1+z_2+\alpha\}$, as well as from  sectors $\beta=\{S+\alpha,S+z_1+\alpha\}$ upon the action of a right-moving fermion oscillator. The latter states are of the form:
\begin{align}
\left|(\psi^{\mu},\chi^{12,34,56})\right>_L\times
\left(\bar{\phi}_{-\frac{1}{2}}
\atop
\bar{\phi}^{\ast}_{-\frac{1}{2}}
\right)
\left|\left(\bar{\psi}^{4,5},\bar{\phi}^{1,2}\right)\right>_R,\;\bar{\phi}_{-1/2}\in\{\bar{\eta}^{1,2,3},\bar{\phi}^{3,\dots,8}\}\nonumber
\end{align}
and
\begin{align}
\left|(\psi^{\mu},\chi^{12,34,56})\right>_L\times
\left(\bar{\phi}_{-\frac{1}{2}}
\atop
\bar{\phi}^{\ast}_{-\frac{1}{2}}
\right)
\left|\left(\bar{\psi}^{4,5},\bar{\phi}^{3,4}\right)\right>_R,\;\bar{\phi}_{-1/2}\in\{\bar{\eta}^{1,2,3},\bar{\phi}^{1,2},\bar{\phi}^{5,\dots,8}\}\nonumber
\end{align}
respectively. Furthermore, acting on the Ramond vacuum of sectors $\beta=\{S+\alpha,S+z_1+\alpha\}$ with a right moving $\bar{\psi}^{1,2,3}$-oscillator  generates states transforming as vectorials of $SU(4)$. We do not consider states arising from the action of $\bar{y}$- or $\bar{\omega}$-oscillators to be relevant, as these do not appear in the massless spectrum of the theory in generic points of the moduli space.

The projection operators corresponding to the four sectors $a$, $\beta$ are $\mathds{P}^-_{a}$ and $\mathds{P}^{\bar{\phi}}_{\beta}$, where:
\begin{equation}
\begin{aligned}
&\Xi^-(S+z_2+\alpha)=\{T_1,T_2,T_3,x+z_1+\alpha\} \\
&\Xi^-(S+z_1+z_2+\alpha)=\{T_1,T_2,T_3,x+\alpha\} \\
&\Xi(S+\alpha)=\{T_1,T_2,T_3,z_2,x+z_1+\alpha\} \\
&\Xi(S+z_1+\alpha)=\{T_1,T_2,T_3,z_2,x+\alpha\}\,,
\end{aligned}
\end{equation}
It has been shown that all exotic fractionally charge states can be projected out from the massless spectrum for supersymmetric string models utilising real fermions \cite{Assel:2009xa}. However, this does not apply here because we use solely complex fermions. As a minimum phenomenological requirement,
 we will impose in the sequel that all exotic states are vector-like, so they can become massive. This is translated to $n_4=n_{\bar{4}}$, as all exotic fermionic state representations are real. For this reason we have also omitted the analysis of the bosonic exotic states in the previous discussion.

As was mentioned above, the gauge symmetry is  enhanced for special values of the GGSO coefficients. 
As a matter of fact, 
additional gauge bosons can arise in the sectors  $a\in\{ z_1+z_2,  z_2+\alpha, z_1+z_2+\alpha, x, \alpha+x, z_1+\alpha+x\}$, upon the action of the left-moving space-time fermion oscillator $\psi^\mu_{-1/2}$, and in the sectors  $\beta\in\{z_1,z_2,\alpha,z_1+\alpha\}$ upon the action of $\psi^\mu_{-1/2}$ and 
the action of a right-moving fermion oscillator. These are of the form

\begin{equation}
\psi^{\mu}_{-\frac{1}{2}}\left|0\right>_L\times
\left|({\alpha}_R)\right>_R
\,,\\
\psi^{\mu}_{-\frac{1}{2}}\left|0\right>_L\times
\left(\bar{\phi}_{-\frac{1}{2}}
\atop
\bar{\phi}_{-\frac{1}{2}}^\ast
\right)
\left|(\beta_R)\right>_R, \; 
\bar{\phi}\in \mathds{1}_R-{\beta}_R\,.
\end{equation}
Since our study is focused on models with Pati-Salam gauge symmetry, we must ensure that all gauge bosons that lead to enhancements of the observable gauge group are eliminated from the spectrum. 
The way
to achieve this is to set
\begin{align}
&\mathds{P}_a^+\mathds{P}_a^-=0, \;a\in\{x,\alpha+x,z_1+\alpha+x,z_2+\alpha,z_1+z_2+\alpha\}, \nonumber\\
&\mathds{P}_a^{\bar{\phi}}=0, \;\bar{\phi}\in \mathds{1}_R-{a}_R,  \;a\in\{\alpha,z_1+\alpha\}, \\
&\mathds{P}_a^{\bar{\phi}}=0, \;\bar{\phi}\in \mathds{1}_R-{a}_R, \; \bar{\phi}\notin\{\bar{\eta}^{1,2,3},\bar{\phi}^{1,\dots,8}\},  \;a\in\{z_1,z_2\}, \nonumber
\end{align}	
where:
\begin{align}
&\Xi^+(x)=\{T_1,T_2,T_3,z_1,z_2\}\,,\nonumber\\
&\Xi^+(\alpha+x)=\{T_1,T_2,T_3,z_2,z_1+\alpha\}
\,,
\nonumber\\
&\Xi^+(z_1+\alpha+x)=\{T_1,T_2,T_3,z_2,\alpha\}
\,, \nonumber\\
&\Xi^+(z_2+\alpha)=\{T_1,T_2,T_3,b_1+b_2,b_1+z_1+\alpha\}
\,,\nonumber\\
&\Xi^+(z_1+z_2+\alpha)=\{T_1,T_2,T_3,b_1+b_2,b_1+\alpha\}\,,\\
&\Xi^-(a)=\{S\}\, ,a\in\{x,\alpha+x,z_1+\alpha+x,z_2+\alpha,z_1+z_2+\alpha\}\,,\nonumber\\
&\Xi(\alpha) = \left\{S,T_1,T_2,T_3,b_1+b_2,b_1+z_1+x,z_2\right\} \,,\nonumber\\
&\Xi(z_1+\alpha) = \left\{S,T_1,T_2,T_3,b_1+b_2,b_1+\alpha,z_2\right\} \,,\nonumber\\
&\Xi(z_1) = \left\{S,T_1,T_2,T_3,b_1,b_2,z_2\right\}\,, \nonumber\\
&\Xi(z_2) = \left\{S,T_1,T_2,T_3,b_1,b_2,z_1,\alpha\right\}\,.  \nonumber
\end{align}
Nevertheless, we will allow for pure hidden sector gauge group enhancements. These include  $z_1$ and $z_2$ sector gauge bosons comprised of oscillators of $\bar{\eta}^{1,2,3}$ or $\bar{\phi}^{1,\dots,8}$.
 More specifically, in the $z_1$ sector, the states
\begin{align}
\psi^{\mu}_{-\frac{1}{2}}\left|0\right>_L\times
\left(\bar{\eta}^{1,2,3}_{-\frac{1}{2}}
\atop
\bar{\eta}^{\ast1,2,3}_{-\frac{1}{2}}
\right)
\left|\left(\bar{\phi}^{1,\dots,4}\right)\right>_R\ \nonumber
\end{align}
which appear in the spectrum when $\mathds{P}_{z_1}^{\bar{\varphi}}=1$, $\bar{\varphi}\in\{{\bar{\eta}^{1,2,3}}\}$,
lead to the  gauge symmetry enhancement $SU(2)_{1/2}\times SU(2)_{3/4}\times U(1)\to SO(6)$, while
\begin{align}
\psi^{\mu}_{-\frac{1}{2}}\left|0\right>_L\times
\left(\bar{\phi}^{5,\dots,8}_{-\frac{1}{2}}
\atop
\bar{\phi}^{\ast5,\dots,8}_{-\frac{1}{2}}
\right)
\left|\left(\bar{\phi}^{1,\dots,4}\right)\right>_R\ \nonumber
\end{align}
enlarge $SU(2)_{1/2}\times SU(2)_{3/4}\times SO(8)$ to $SO(12)$ and are present when $\mathds{P}_{z_1}^{\bar{\varphi}}=1$, $\bar{\varphi}\in\{{\bar{\phi}^{5,\dots,8}}\}$. Likewise, in $z_2$, the states
\begin{align}
\psi^{\mu}_{-\frac{1}{2}}\left|0\right>_L\times
\left(\bar{\eta}^{1,2,3}_{-\frac{1}{2}}
\atop
\bar{\eta}^{\ast1,2,3}_{-\frac{1}{2}}
\right)
\left|\left(\bar{\phi}^{5,\dots,8}\right)\right>_R\ \nonumber
\end{align}
result in $U(1)\times SO(8)\to SO(10)$, while
\begin{align}
\psi^{\mu}_{-\frac{1}{2}}\left|0\right>_L\times
\left(\bar{\phi}^{1,2}_{-\frac{1}{2}}
\atop
\bar{\phi}^{\ast1,2}_{-\frac{1}{2}}
\right)
\left|\left(\bar{\phi}^{5,\dots,8}\right)\right>_R\, \text{and}\,  \psi^{\mu}_{-\frac{1}{2}}\left|0\right>_L\times
\left(\bar{\phi}^{3,4}_{-\frac{1}{2}}
\atop
\bar{\phi}^{\ast3,4}_{-\frac{1}{2}}
\right)
\left|\left(\bar{\phi}^{5,\dots,8}\right)\right>_R \nonumber
\end{align}
are responsible for  the enlargements $SU(2)_1\times SU(2)_2\times SO(8)\to SO(12)$ and $SU(2)_3\times SU(2)_4\times SO(8)\to SO(12)$ respectively.
Finally, gauge bosons arising in the $z_1+z_2$ sector result in the enhancement of the hidden $SU(2)_{1/2}\times SU(2)_{3/4}\times SO(8)$ gauge symmetry to $SO(12)$. The projection operator signaling their presence is:
	\begin{align}
	\mathds{P}_{z_1+z_2}^{+}\mathds{P}_{z_1+z_2}^{-}=1, \ \text{with}\ 
	\Xi^+(z_1+z_2)=\{T_1,T_2,T_3,b_1,b_2\}\ ,\ 
	\Xi^-(z_1+z_2)=\{S\}\,.
	\end{align}
	Since the observable Pati--Salam gauge symmetry is unaffected when such enhancements take place, the corresponding models are relevant to our analysis. It turns out that after imposing all phenomenological constraints introduced in Section 2, ensuring SUSY is spontaneously broken \`{a}-la Scherk--Schwarz and identifying cases in which the effective one-loop potential is of the super no-scale type, only four possible enhancements of the hidden sector remain:
	$U(1)^3\times SU(2)^2\times SO(12)$, $U(1)^2\times SU(2)^2\times SO(6)\times SO(8)$, $U(1)^2\times SU(2)^4\times SO(10)$ and $U(1)^2\times SO(6)\times SO(12)$.

\section{Orbifolds, SUSY Breaking, and the Cosmological Constant \label{section3}}

Although providing a powerful tool for string model building and for analysing the corresponding spectra, the fermionic formulation of superstring theory essentially provides a ``snapshot'' of the theory valid only at a specific ``fermionic point" of the perturbative moduli space. Indeed, the two dimensional CFT of each pair of real world-sheet fermions $\{y^I(z),\omega^I(z)\}$ may be consistently replaced by that of a chiral boson at fixed radius $R_I=\sqrt{\alpha'/2}$, such that $i\partial X^I(z) = {:y^I(z)\omega^I(z):}$, and similarly for the right movers. The models constructed thus far in the free fermionic framework with purely real boundary conditions are hence mapped to compactifications on a $T^6/\mathbb Z_2^N$ orbifold involving $\mathbb Z_2$ rotations $r_I: \{X^I \to -X^I\}$ of the internal world-sheet (super)coordinates, potentially supplemented by order-two shifts $\delta_I:\{X^I\to X^I+\pi \sqrt{\alpha'/2}\}$. The sigma model of the theory may then be consistently deformed away from the fermionic point by turning on marginal deformations of the current-current type, controlled by the VEVs of the no-scale moduli.

Contrary to the case of string vacua with unbroken spacetime supersymmetry, it is clear that, in non-supersymmetric constructions, an analysis based purely on fermionic point data no longer suffices for a consistent study of the theory. We can outline two immediate issues. Firstly, as discussed in the introduction, the absence of supersymmetry is already accompanied by the generation of a non-trivial scalar potential at the one loop order. Depending on its shape, the no-scale moduli are then typically driven to other regions of moduli space, where the masses of some towers of states will be shifted and the massless spectrum of the theory will look very different from the one extracted at the fermionic point. 

Secondly, at the multi-critical fermionic point, rotations $r_I$ (twists) and translations $\delta_I$ (shifts) are effectively treated at the same level, encoded as boundary conditions of the fermionic world-sheet fields and are, at first sight, indistinguishable. As a result, questions involving  moduli dependence and, importantly, the type of supersymmetry breaking in play are obscured when working at the fermionic point. It is only after the map to the orbifold formulation has been performed, and the moduli dependence restored, that one may consistently ascertain whether spacetime supersymmetry is broken spontaneously \`a la Scherk-Schwarz, or explicitly.

In what follows, we explicitly restore the no-scale moduli dependence and further explore the perturbative parameter space of the models by mapping them to orbifold compactifications using the procedure outlined in \cite{Florakis:2016ani}.  The compactification relevant for our case is based on the symmetric $T^6/\mathbb Z_2\times \mathbb Z_2$ orbifolds of the $E_8\times E_8$ heterotic string, where the internal 6-torus is split into a product of three $2$-tori. The action of the $\mathbb Z_2\times\mathbb Z_2$ orbifold preserves $\mathcal N_4=1$ supersymmetry and corresponds to the singular limit of a Calabi-Yau, while the standard embedding on the gauge bundle side corresponds to breaking one of the $E_8$ gauge group factors down to $SO(10)\times U(1)^3$. Freely-acting $\mathbb Z_2$ factors are then introduced in order to realise the spontaneous breaking of supersymmetry \`a la Scherk-Schwarz. This is achieved by correlating the U(1) R-symmetry charges of string states to monodromies along non-trivial cycles of the compactification 2-tori. Finally, additional $\mathbb Z_2$ factors are needed to implement the reduction to the Pati-Salam gauge group factors \eqref{Gauge Symmetry}. 

The one-loop partition function can be expressed in terms of Dedekind $\eta(\tau)$ and Jacobi theta constants with characteristics $\vartheta[^a_b](\tau)$, in addition to the product of the three Lorentzian (2,2) lattices parametrising the 2-tori:
\begin{equation}\label{Orbifold PF}
\begin{aligned}
Z&=\frac{1}{\eta^{2}\bar{\eta}^{2}}\ \frac{1}{2^4}\sum_{\substack{h_1,h_2,H,H'\\g_1,g_2,G,G'}}\frac{1}{2^3}\sum_{\substack{a,k,\rho\\b,\ell,\sigma}}\frac{1}{2^3}\sum_{\substack{H_1,H_2,H_3\\G_1,G_2,G_3}}(-1)^{a+b+HG+H'G'+\Phi} \\
&\times \frac{\vartheta[^a_b]}{\eta}\,\frac{\vartheta[^{a+h_1}_{b+g_1}]}{\eta}\,\frac{\vartheta[^{a+h_2}_{b+g_2}]}{\eta}\,\frac{\vartheta[^{a-h_1-h_2}_{b-g_1-g_2}]}{\eta}\, \\
&\times \frac{\bar{\vartheta}[^{k}_{\ell}]^3}{\bar\eta^3}\,\frac{\bar{\vartheta}[^{k+H'}_{\ell+G'}]}{\bar\eta}\,\frac{\bar{\vartheta}[^{k-H'}_{\ell-G'}]}{\bar\eta}\,
\frac{\bar{\vartheta}[^{k+h_1}_{\ell+g_1}]}{\bar\eta}\,\frac{\bar{\vartheta}[^{k+h_2}_{\ell+g_2}]}{\bar\eta}\,\frac{\bar{\vartheta}[^{k-h_1-h_2}_{\ell-g_1-g_2}]}{\bar\eta}\, \\
&\times \frac{\bar{\vartheta}[^{\rho+H'}_{\sigma+G'}]}{\bar\eta}\,\frac{\bar{\vartheta}[^{\rho-H'}_{\sigma-G'}]}{\bar\eta}\,\frac{\bar{\vartheta}[^{\rho}_{\sigma}]^2}{\bar\eta^2}\,\frac{\bar{\vartheta}[^{\rho+H}_{\sigma+G}]^4}{\bar\eta^4}\\
&\times  \frac{\Gamma^{(1)}_{2,2}[^{H_1}_{G_1}|^{h_1}_{g_1}](T^{(1)},U^{(1)})}{\eta^2\bar\eta^2}\ \frac{\Gamma^{(2)}_{2,2}[^{H_2}_{G_2}|^{h_2}_{g_2}](T^{(2)},U^{(2)})}{\eta^2\bar\eta^2}\ \frac{\Gamma^{(3)}_{2,2}[^{H_3}_{G_3}|^{h_1+h_2}_{g_1+g_2}](T^{(3)},U^{(3)})}{\eta^2\bar\eta^2} \,.
\end{aligned}
\end{equation}
All summations appearing in this expression are understood to be in $\mathbb Z_2$. For the left-moving RNS fermions, $a=0$ and $a=1$ label the Neveu-Schwarz (NS) and Ramond (R) sectors, respectively, whereas the $b$-summation in the presence of the cocycle phase $(-1)^{b}$ imposes the standard GSO projection. Similarly, the pairs $(k,\ell)$ and $(\rho,\sigma)$ label the boundary conditions of the $16$ complex fermions realising the level-one Kac-Moody lattice for the right-movers.

The pair $(h_1,h_2)$ labels the twisted sectors of the non-freely acting $\mathbb Z_2\times\mathbb Z_2$ orbifold group corresponding to the CY, while the projection onto invariant states is achieved by the sum over $(g_1,g_2)$. Three additional freely acting $\mathbb{Z}_2$ orbifolds involving order-two shifts on each of the three $T^2$ tori are similarly parametrised by $(H_i,G_i)$ with ${i=1,2,3}$, and correspond to the Scherk-Schwarz breaking. Finally, the pairs $(H,G)$ and $(H',G')$ are responsible for further twisting the Kac-Moody currents and breaking the non-abelian gauge symmetry to the observable Pati-Salam group times the factor $SO(4)^2\times SO(8)$ from the hidden sector.

The moduli dependence of the theory is made manifest via the twisted/shifted $(2,2)$ toroidal lattices given by:
\begin{equation}\label{lattice definition}
\Gamma^{(i)}_{2,2}[^{H_i}_{G_i}|^{h_i}_{g_i}](T^{(i)},U^{(i)})=\left\{
\begin{array}{ll}
\abs{\frac{2\eta^3}{\vartheta\begin{bsmallmatrix}
	1-h_i \\ 1-g_i 
	\end{bsmallmatrix}}}^2\text{ , } & 
\begin{smallmatrix}
(H_i,G_i)=(0,0) & \text{ or } \\
\;\;\;(H_i,G_i)=(h_i,g_i) & \\
\end{smallmatrix} \\
\\
\Gamma_{2,2}^{\text{shift}}\begin{bsmallmatrix}
H_i \\ G_i 
\end{bsmallmatrix}(T^{(i)},U^{(i)})\text{ , } & (h_i,g_i)=(0,0) \\
\\
0 \text{ , } & \text{otherwise}
\end{array} 
\right. 
\end{equation}
The shifted lattice is given as the sum:
\begin{equation}\label{shifted lattice}
\Gamma_{2,2}^{\text{shift}}\begin{bsmallmatrix}
H_i \\ G_i 
\end{bsmallmatrix}(T^{(i)},U^{(i)})=\sum_{\substack{m_1,m_2\\n_1,n_2}\in\mathbb{Z}}(-1)^{G(m_1+n_2)}q^{\abs{p_L}^2/4}\bar{q}^{\abs{p_R}^2/4},
\end{equation}
over the lattice momenta:
\begin{equation}
\begin{aligned}
p_L=\frac{m_2+\frac{H_i}{2}-U^{(i)}m_1+T^{(i)}(n_1+\frac{H_i}{2}+U^{(i)}n_2)}{\sqrt{T^{(i)}_2U^{(i)}_2}} \\
p_R=\frac{m_2+\frac{H_i}{2}-U^{(i)}m_1+\overline{T}^{(i)}(n_1+\frac{H_i}{2}+U^{(i)}n_2)}{\sqrt{T^{(i)}_2U^{(i)}_2}}
\end{aligned}
\end{equation}
where $T^{(i)}_2$, $U^{(i)}_2$ denote the imaginary parts of the K\"ahler and complex structure moduli $T^{(i)}$ and $U^{(i)}$, respectively.

The correspondence with the free fermionic formulation is obtained by fixing the moduli of the three 2-tori to their fermionic point values: $T^{(1,2,3)}=i$ and $U^{(1,2,3)}=(1+i)/2$, in which case the lattices factorise into products of Jacobi constants as in \cite{Florakis:2016ani}
\begin{equation}\label{fermion-lattice equation}
\Gamma_{2,2}\begin{bsmallmatrix}
H_i & h_i \\ G_i & g_i
\end{bsmallmatrix}\big(i,(1+i)/2)\big)=\frac{1}{2}\sum_{\epsilon,\zeta=0,1}\abs{\vartheta\begin{bsmallmatrix}
\epsilon \\ \zeta
\end{bsmallmatrix}\vartheta\begin{bsmallmatrix}
\epsilon+h_i \\ \zeta+g_i
\end{bsmallmatrix}}^2(-1)^{H_i(\zeta+g_i)+G_i(\epsilon+h_i)+H_iG_i}.
\end{equation}
The modular invariant phase $\Phi$, constructed out of boundary condition bilinears, corresponds to the GGSO projections of the fermionic construction defined by the choice of $c[^{\beta_i}_{\beta_j}]$. A special choice of this phase is responsible for implementing the Scherk-Schwarz breaking by correlating the R-symmetry charges to the $T^2$ lattice momenta. Although several choices are a priori possible, for technical convenience, it will be sufficient to couple the R-symmetry charges only to the first 2-torus. With this choice, the gravitino acquires moduli dependent mass $m_{3/2}= |U|/\sqrt{T_2 U_2}$ parametrised by the no-scale moduli of the Scherk-Schwarz 2-torus.

The one-loop effective potential can be obtained by computing the vacuum amplitude of the theory as a function of the VEVs of the non-scale moduli $\{t_I\}$. This essentially involves the integration 
\begin{equation}\label{effective potential}
V_{\text{one-loop}}(t_I)=-\frac{1}{2(2\pi)^4}\int_{\mathcal{F}}\frac{d^2\tau}{(\Im\tau)^3}Z(\tau,\bar{\tau};t_I),
\end{equation}
of the string partition function over the moduli space $\mathcal{F}=SL(2;\mathbb{Z})\backslash \mathbb{H}^+$ of complex structures $\tau$ of the world-sheet torus. Whenever the theory preserves at least one unbroken target space supersymmetry, the integral  vanishes identically as a result of the exact bose-fermi degeneracy present already in the partition function at all mass levels, which originates from the spectral flow of a global $\mathcal N_2=(2,2)$ SCFT on the string world sheet.

Different is the case of non-supersymmetric constructions, where the effective potential is no longer superprotected, the bose-fermi degeneracy of the string spectrum is disrupted and the vacuum amplitude receives different contributions from the full towers of bosonic and fermionic states in the string spectrum. In general, the computation of the one-loop potential in this case is a highly non-trivial technical problem, involving the integration of a weak, non-holomorphic modular function over the fundamental domain $\mathcal F$. Some recent developments of new modular integration techniques and associated discussions of the related convergence difficulties can be found in \cite{Angelantonj:2011br,Angelantonj:2012gw,Angelantonj:2013eja,Angelantonj:2015rxa,Florakis:2016boz}.

By means of the theta lift of eq. \eqref{effective potential}, the integral inherits the T-duality symmetry of the theory, encoded in the moduli-dependent (shifted) lattices in $Z(\tau,\bar\tau;t_I)$. As a result, the one-loop potential becomes an automorphic form of the residual T-duality group $\prod O(2,2;\mathbb Z)_{T^i,U^i}/\sigma$ left unbroken by the freely acting orbifold translations parametrised by $(H_i,G_i)$. To date, there is no fully satisfactory analytic method known in the literature for treating such integrals in a way that preserves the full T-duality symmetry manifest. In fact, in the case of BPS protected couplings, the quasi-holomorphic structure of the integrand allows the application of powerful new techniques based on Poincar\'e series that allow the integration to be performed while manifestly preserving the T-duality symmetry. Unfortunately, those techniques are difficult to apply in the non-BPS amplitude \eqref{effective potential}, due to the lack of holomorphy as well as convergence issues that arise near self-dual points of the perturbative moduli space (for a discussion, see \cite{Florakis:2016ani}). Therefore, the unambiguous way to study the effective potential close to self dual points $t_I\sim1$ (in string units), is by employing numerical integration methods.

Additional insight can be extracted by considering large volume regions in the internal space. In particular, as long as the characteristic length scale of the compactification manifold is sufficiently larger than the fermionic radius, the modular integral of eq.\eqref{effective potential} can be consistently computed by lattice unfolding methods \cite{Dixon:1990pc}. For technical convenience, we keep the moduli of the spectator 2-tori at the fermionic point, and deform only the moduli of the first $T^2$ that directly control the Scherk-Schwarz breaking. 

The summation over the spin-structures in eq.\eqref{Orbifold PF} may be performed using the Riemann-Jacobi identity. Notice that the sector $H_1=G_1=0$ is the original supersymmetric theory, since it amounts to ``switching off'' the Scherk-Schwarz boost and, hence, its contribution to the partition function vanishes due to the usual bose-fermi degeneracy at all mass levels. Supersymmetry is also unbroken in the sectors $(H_1,G_1)=(h_1,g_1)$, corresponding to the spontaneous breaking of $\mathcal{N}=2\rightarrow1$. In all other cases, the twisted/shifted lattice of eq.\eqref{lattice definition} vanishes, unless $(h_1,g_1)=(0,0)$. Therefore, only the sector $h_1=g_1=0$ contributes to the one-loop potential. We then organise the partition function eq.\eqref{Orbifold PF} into orbits of the freely acting $\mathbb Z_2$ responsible for the spontaneous breaking of supersymmetry  
\begin{equation}\label{BlocksScherkSchwarz}
Z=\frac{1}{2}\sum_{H_1,G_1\in\mathbb Z_2}\Psi[^{H_1}_{G_1}]\,\Gamma_{2,2}^{\text{shift}}[^{H_1}_{G_1}](T,U) \,,
\end{equation}
and Fourier expand the moduli independent blocks in each orbifold sector
\begin{equation}
	\Psi[^{H_1}_{G_1}] = \sum_{\Delta\geq -\frac{1}{2}}\sum_{\bar \Delta\geq -1} c[^{H_1}_{G_1}](\Delta,\bar\Delta)\,q^\Delta \,\bar q^{\bar\Delta} \,,
\end{equation}
in terms of the nome $q=e^{2\pi i\tau}$. Here, $c[^{H_1}_{G_1}](\Delta,\bar\Delta)$ counts the number of bosonic minus fermionic states with left-right conformal weights $(\Delta,\bar\Delta)$ in the $(H_1,G_1)$ sector.

After carefully unfolding the fundamental domain using the orbits of the (2,2) shifted lattice as in \cite{Florakis:2016ani}, the dominant contribution\footnote{An additional contribution from the non-degenerate orbit of the shifted (2,2) lattice can be seen to decay exponentially faster than the second line of \eqref{DegenerateOrbit} and we do not display it explicitly here.
} to the one-loop potential valid at large volume ${\rm vol}(T^2)_1 = T_2\gg 1$ takes the form
\begin{equation}\label{DegenerateOrbit}
\begin{split}
	2(2\pi)^4 V_{\text{one-loop}}(T,U) &\simeq -\frac{c[^0_1](0,0)}{\pi^3 T_2^2} \sum_{m_1,m_2\in\mathbb Z}\frac{U_2^3}{|m_1+\frac{1}{2}+Um_2|^6}\\
	-\frac{2\sqrt{2}}{\sqrt{T_2}}&\sum_{\Delta>0} \Delta^{3/2} c[^0_1](\Delta,\Delta) \sum_{m_1,m_2\in\mathbb Z}\frac{U_2^{3/2}}{|m_1+\frac{1}{2}+Um_2|^3} \,K_3\left(2\pi\sqrt{\frac{\Delta T_2}{U_2}}\left|m_1+\tfrac{1}{2}+Um_2\right|\right)\,.
\end{split}
\end{equation}
The first line corresponds to the leading behaviour at large volume and reproduces the power law suppression of \cite{Antoniadis:1990ew}. Notably, it is controlled by the difference $\lambda_0 \equiv \frac{1}{2}c[^0_1](0,0)$ in the number of bosons and fermions that remain massless at the \emph{generic point} of the deformation in the $(T,U)$ moduli space.

Super no-scale models \cite{Harvey:1998rc,Kounnas:2016gmz,Angelantonj:1999gm,Shiu:1998he} correspond to theories where, despite the absence of spacetime supersymmetry, the massless spectrum still enjoys a bose-fermi degeneracy. In our case, where supersymmetry is broken spontaneously by a freely-acting orbifold, this amounts to satisfying the condition $\lambda_0=0$, with drastic consequences for the structure of the one-loop potential.

It is actually possible to understand the basic features of the one-loop potential in super no-scale models and the quantities that control its shape in a model independent fashion. Clearly, in the infinite volume limit $T_2\to\infty$, the gravitino becomes massless and supersymmetry is restored. The potential, therefore, must vanish in this limit and this behavior is precisely reflected by the above asymptotic expression. Imposing the super no-scale condition then removes the power law term and the one-loop potential becomes exponentially suppressed at large volume, due to the Bessel function in the second line of \eqref{DegenerateOrbit}. 

Whether the potential asymptotically approaches its vanishing limit from positive or negative values is dictated by the first non-vanishing degeneracy in first mass levels (to be computed after factoring out the Scherk--Schwarz lattice). If we denote their conformal weight as\footnote{In the models we construct, $\Delta_{\min}=1/4$.} $\Delta_{\min}>0$, the sign of the potential is then simply given by minus the sign of $\lambda_{1} \equiv \frac{1}{2}c[^0_1](\Delta_{\min},\Delta_{\min})$.  If $\lambda_1>0$, by argument of continuity, the minimum of the potential will necessarily lie at large negative values $V_{\min}<0$ and close to the self-dual points $T_2\sim 1$ under the residual T-duality group of the theory, where the potential is forced to display extrema. Around the self-dual points the potential is no longer exponentially suppressed and \eqref{DegenerateOrbit} is no longer valid due to convergence issues. Requiring $\lambda_{1}<0$ is, therefore, a necessary condition for dynamical decompactification, which should be imposed together with the super no-scale condition $\lambda_0=0$, and ensures the exponential suppression of the cosmological constant at large volume.

Additional information about the behaviour of the potential at radii of the order of the string scale can be extracted by numerically integrating \eqref{effective potential} at the fermionic point. For example, choosing a vanishing B-field background and keeping the complex structure modulus fixed at the fermionic point, a deformation in the $T^2$ volume implies that the potential exhibits the symmetry $V_{\rm one-loop}(T_2)=V_{\rm one-loop}(1/T_2)$ with the fixed point corresponding precisely to the fermionic point $T_2=1$. Depending on the value of the potential at the fermionic point ${V_F} \equiv V_{\rm one-loop}(T_2=1)$,  the extremum exhibited by the potential at this point may be a global maximum (${V_F}>0$), a local minimum (${V_F}>0$), or a global minimum (${V_F}<0$). Taking these considerations into account, we see that consistent super no-scale models are obtained by simultaneously requiring $\lambda_0 =0$, $\lambda_1<0$ and ${V_F} >0$.  In later sections, we impose $\lambda_0=0$ from the start, and classify the  potentials for all models according to the values of $\lambda_1$ and ${V_F}$. We show there that the potentials obtained by numerical integration for all models in our exhaustive search can indeed be classifid according to these parameters.

Before closing this section, we wish to address the question of whether the massless bose-fermi degeneracy $\lambda_0=0$ imposed at the \emph{generic} point of the Scherk-Schwarz 2-torus, could in principle persist also at the fermionic point. It is possible to prove that this can never occur in the models under consideration in this work. The argument, however, has interesting implications for potential generalisations of our construction. First, decompose the blocks of \eqref{BlocksScherkSchwarz} into eigenstates with respect to the Scherk-Schwarz orbifold
\begin{equation}
\begin{split}
	&\Gamma_{2,2}^{\rm shift}[^{H_1}_{\,\,\pm}] = \frac{1}{2}\left( \Gamma_{2,2}^{\rm shift}[^{H_1}_{~0}] \pm \Gamma_{2,2}^{\rm shift}[^{H_1}_{~1}] \right)\,, \\
	&\Psi[^{H_1}_{\,\,\pm}] =\frac{1}{2}\left( \Psi[^{H_1}_{~0}] \pm \Psi[^{H_1}_{~1}]  \right)\,,\\
	& c[^{H_1}_{\,\,\pm}](\Delta,\bar\Delta) = \frac{1}{2}\left( c[^{H_1}_{~0}](\Delta,\bar\Delta) \pm c[^{H_1}_{~1}](\Delta,\bar\Delta)\right)\,,
\end{split}
\end{equation}
and expand the partition function as
\begin{equation}\label{Zdecompose}
	Z = \Psi[^{\,0}_{+}] \,\Gamma_{2,2}^{\rm shift}[^{\,0}_{+}] + \Psi[^{\,0}_{-}] \,\Gamma_{2,2}^{\rm shift}[^{\,0}_{-}] + \Psi[^{\,1}_{+}] \,\Gamma_{2,2}^{\rm shift}[^{\,1}_{+}] + \Psi[^{\,1}_{-}] \,\Gamma_{2,2}^{\rm shift}[^{\,1}_{-}] \,.
\end{equation}
Since $\Psi[^0_0]=0$ from the original unbroken supersymmetry, we also have $\Psi[^{\,0}_{\pm}] = \pm \frac{1}{2}\Psi[^0_1]$. At the generic point in the $(T,U)$ moduli space, the shifted lattice $\Gamma_{2,2}^{\rm shift}[^{\,0}_{+}]= 1+\ldots$ always contains the $\mathbb Z_2$ invariant state with vanishing lattice charges corresponding to the unit operator. Together with the contributions of conformal weight $(0,0)$ from $\Psi[^{\,0}_{+}]$, this constructs bosonic and fermionic states that remain massless at the generic point and their contribution to the partition function is clearly equal to $c[^{\,0}_{+}](0,0) = \lambda_0$. No other massless states can arise at the generic point, since in all other sectors of the decomposition \eqref{Zdecompose}, the shifted lattice always contributes non-trivially and the resulting states carry moduli dependent masses.

On the other hand, at the fermionic point the shifted lattice contributions have the Fourier expansion
\begin{equation}
\begin{split}
	&\Gamma_{2,2}^{\rm shift}[^{\,0}_{+}]= 1+ 16 q^{1/2} \bar q^{1/2} + 4\bar q +\ldots \\
	&\Gamma_{2,2}^{\rm shift}[^{\,0}_{-}]= 8 q^{1/4} \bar q^{1/4}  +\ldots \\
	&\Gamma_{2,2}^{\rm shift}[^{\,1}_{+}]= 8 q^{1/4} \bar q^{1/4} +\ldots \\
	&\Gamma_{2,2}^{\rm shift}[^{\,1}_{-}]=  4q^{1/2}+4\bar q^{1/2} + 16 q^{1/2}\bar q +\ldots 
\end{split}
\end{equation}
where the ellipses stand for terms that are already massive by the lattice alone, regardless of their coupling to $\Psi$. We can now identify the extra massless states. First, notice that aside from the universal unit contribution, level matched contributions of the type $(q\bar q)^{|\Delta|}$ in the untwisted $H_1=0$ lattice never form extra massless states. Indeed, the only way this could happen is if they originated from states in $\Psi[^{0}_1]$ with conformal weight $(\Delta,\Delta)$ and $\Delta= -1/2$ or $\Delta = -1/4$. However, these would then also couple with the unit contribution in $\Gamma_{2,2}[^{\,0}_{+}]$ to form physical tachyons at any point in moduli space, which is clearly not the case for a theory with spontaneous supersymmetry breaking. Moreover, counting the conformal weights from states in $\Psi[^{\,1}_{+}]$ one can show that no extra massless states arise from this sector either. The 16 states of weight $(1/2,1)$ in $\Gamma_{2,2}^{\rm shift}[^{\,1}_{-}]$ similarly give rise to no extra massless states, since this would require coupling to the weight $(-1/2,-1)$ NS ground state, but this is always even under the Scherk-Schwarz orbifold and, hence, does not appear in $\Psi[^{\,1}_{-}]$.

In the untwisted sector, extra massless states can arise from the asymmetric contribution $4\bar q$ of $\Gamma_{2,2}^{\rm shift}[^{\,0}_{+}]$, by coupling it together with the universal protograviton term $2/\bar q$ of $\Psi[^{\,0}_{+}]$. In the twisted sector, the asymmetric contributions $4q^{1/2}$ and $4\bar q^{1/2}$ can generate massless states by coupling with terms in $\Psi[^{\,1}_{-}]$ of conformal weight $(-\frac{1}{2},0)$ and $(0,-\frac{1}{2})$, respectively. Taking into account these multiplicities, and denoting $d_{-\frac{1}{2},0}\equiv c[^{\,1}_{-}](-\tfrac{1}{2},0)$ and $d_{0,-\frac{1}{2}}\equiv c[^{\,1}_{-}](0,-\tfrac{1}{2})$, the total number of massless bosonic minus fermionic states at the fermionic point reads
\begin{equation}
	 \lambda_0+4\times(2+ d_{-\frac{1}{2},0}+d_{0,-\frac{1}{2}} )\,.
\end{equation}
It is now clear that, super no-scale models may only exhibit the massless bose-fermi degeneracy also at the fermionic point  if
\begin{equation}\label{FFcondition}
	2+ d_{-\frac{1}{2},0}+d_{0,-\frac{1}{2}} =0 \,.
\end{equation}
Using an explicit representation for $\Psi$ corresponding to our constructions
\begin{equation}\label{Simplified PF}
\begin{aligned}
\Psi[^{H_1}_{G_1}]&=\frac{1}{\eta^{12}\bar{\eta}^{24}}\frac{1}{2^7}\sum_{\substack{h_2,H,H'\in\mathbb Z_2\\g_2,G,G'\in\mathbb Z_2}}\;\sum_{\substack{k,\rho,\gamma_2,\gamma_3\in\mathbb Z_2\\\ell,\sigma,\delta_2,\delta_3\in\mathbb Z_2}}(-1)^{HG+H'G'+\hat{\Phi}} \\
&\times\vartheta[^{1+H_1}_{1+G_1}]^2\vartheta[^{1+H_1+h_2}_{1+G_1+g_2}]^2 \,\bar{\vartheta}[^{k}_{\ell}]^4\,\bar{\vartheta}[^{k+H'}_{\ell+G'}]\,\bar{\vartheta}[^{k-H'}_{\ell-G'}]\bar{\vartheta}[^{k+h_2}_{\ell+g_2}]\bar{\vartheta}[^{k-h_2}_{\ell-g_2}],\\
&\times\bar{\vartheta}[^{\rho+H'}_{\sigma+G'}]\,\bar{\vartheta}[^{\rho-H'}_{\sigma-G'}]\,\bar{\vartheta}[^{\rho}_{\sigma}]^2\,\bar{\vartheta}[^{\rho+H}_{\sigma+G}]^4 \abs{\vartheta[^{\gamma_2}_{\delta_2}]\vartheta[^{\gamma_2+h_2}_{\delta_2+g_2}]}^2\abs{\vartheta[^{\gamma_3}_{\delta_3}]\vartheta[^{\gamma_3-h_2}_{\delta_3-g_2}]}^2 \,,
\end{aligned}
\end{equation}
and exploiting the periodicity properties of Jacobi theta constants to group them into squares, it is straightforward to see that $d_{0,-\frac{1}{2}}$ and $d_{-\frac{1}{2},0}$ are both $0\,({\rm mod}\, 4)$ and, therefore, \eqref{FFcondition} is always violated. 

This explains the fact that the super no-scale models we construct always exhibit a mismatch in the number of bosons and fermions that are massless at the fermionic point. By an extension of this analysis, a similar argument could be used in order to derive the condition analogous to \eqref{FFcondition} for other constructions, e.g. in models with different shift action on the toroidal lattices. As a result, preserving the super no-scale property also at special points of the perturbative moduli space is far from obvious. The argument can be turned around, to indicate that theories with massless bose-fermi degeneracy at the fermionic point typically lose this property as soon as they are deformed in moduli space, since additional conditions \eqref{FFcondition} would then be needed in order to ensure that the net contribution of the extra massless states vanishes separately and the super no-scale condition $\lambda_0=0$ is satisfied. It should be stressed, as outlined also in the introduction, that the super no-scale property $\lambda_0=0$ is desirable at the generic point since it directly controls the exponential suppression of the cosmological constant. Typically, having massless bose-fermi degeneracy only at the fermionic point in a theory with spontaneously broken supersymmetry, does not lead to exponential suppression in the one-loop potential, since massive and even non-level matched states which still produce sizeable contributions to the modular integral, as explained in \cite{Florakis:2016ani}.


\section{Exploring the models' parameter space\label{section4}}

In Section \ref{section2} we defined the class of non-supersymmetric Pati--Salam models under consideration in the free-fermionic formulation and further mapped it to the corresponding orbifold construction in Section \ref{section3}. Having discussed their salient features, including the absence of tachyon instabilities, the implementation of the Scherk--Schwarz mechanism for the breaking of supersymmetry, the existence of chiral fermion generations as well as the presence of the required gauge symmetry breaking Higgs scalars, we proceed in this section with the comprehensive scan of the space of models and the detailed analysis of their characteristics.

Following the discussion in Section \ref{section2}, our construction gives rise to a huge number of models, approximately $3.5\times10^{13}$, which are \emph{in principle} distinct. However, it turns out that only few of these are compatible with a minimal set of phenomenological requirements. The identification of these Pati--Salam models poses an interesting challenge. To solve it, we put forward a set of criteria that enable us to select and classify models which exhibit a range of desirable features. These are: 
\begin{enumerate}[label=(\alph*)]
\item Absence of physical tachyons in the string spectrum, as defined in \eqref{apt}.
\item Existence of complete chiral fermion generations, as formulated in  \eqref{efga}, \eqref{netchirality}; that is $n_g\ne0$, where $n_g=n_{L}-\bar{n}_{L}= \bar{n}_{R}-n_{R}$.
\item Existence of Pati--Salam and SM symmetry breaking scalar Higgs fields; that is $n_H\ge1, n_h\ge1$, as outlined in Eqs. \eqref{HPS}, \eqref{HSM}.
\item Absence of observable gauge group enhancements, as detailed at the end of Section \ref{section2}.
\item Vector-like fractionally charged exotic states; that is  $n_4=n_{\bar{4}}$, as explained in Section \ref{section2}.
\item Consistency with the  Scherk--Schwarz spontaneous SUSY breaking, as elaborated in Section \ref{section3} and in \cite{Florakis:2016ani}.
\item Compliance with the super-no-scale condition, $\lambda_0=0$, that is translated to equality of the fermionic and bosonic degrees of freedom  at the generic point, as explained in Section \ref{section3}.
\end{enumerate}
For models that satisfy the above requirements, we also
perform a numerical evaluation of the one-loop effective potential as a function of the $T_2$ modulus using \eqref{effective potential} and classify its shape and asymptotic behaviour at large volume $T_2\to\infty$, utilising the values $V_{F}$ and $\lambda_1$, in accordance with the discussion in Section \ref{section3}. Furthermore, we 
deduce the hidden sector gauge group and extract additional details for each model, such as the number of PS and SM breaking Higgs scalar multiplets and the number of fractional charge exotics, as outlined in Section \ref{section2}.

At a technical level, the aforementioned criteria are incorporated into a computer program that performs a full scan over all GGSO projection coefficients $c\left[\beta_i\atop\beta_j\right]$, using  classification techniques developed in \cite{Gregori:1999ny,Faraggi:2004rq,Assel:2010wj}. Although there are \emph{a priori} 45 independent GGSO related phases, this number can be actually reduced down to 34, without loss of generality. Indeed, the four coefficients $c\left[\mathds{1}\atop\mathds{1}\right]$,  $c\left[\mathds{1}\atop S\right]$ and $c\left[S\atop b_a\right]$  correspond to conventions, while $c\left[b_1\atop b_2\right]$ is related to an overall chirality flip. Furthermore, six additional coefficients $c\left[T_1\atop b_2\right]$ and  $c\left[T_2\atop b_1\right]$, $c\left[\mathds{1}\atop z_a\right], c\left[\mathds{1}\atop b_a\right]$  are found to be irrelevant for the implementation of the above criteria. This effectively narrows the search space down to $2^{34} \sim 1.7\times 10^{10}$ configurations. The computer analysis can be further accelerated using a two stage scan \cite{Faraggi:2017cnh,Faraggi:2019qoq}. In the first stage we scan all $SO(10)$ configurations generated by the first nine basis vectors $\beta_1,\dots,\beta_9$ together with the associated GGSO projections, and identify ``fertile" configurations, \emph{i.e.} GGSO phases sets which could in principle lead to acceptable PS models upon the introduction of the additional vector $\beta_{10}=\alpha$. In the second stage, the last vector $\alpha$ and the corresponding GGSO projections are then considered exclusively in conjunction with ``fertile" $SO(10)$ GGSO sets.

\begin{figure}[h!]
\centering
\includegraphics[scale=1.2]{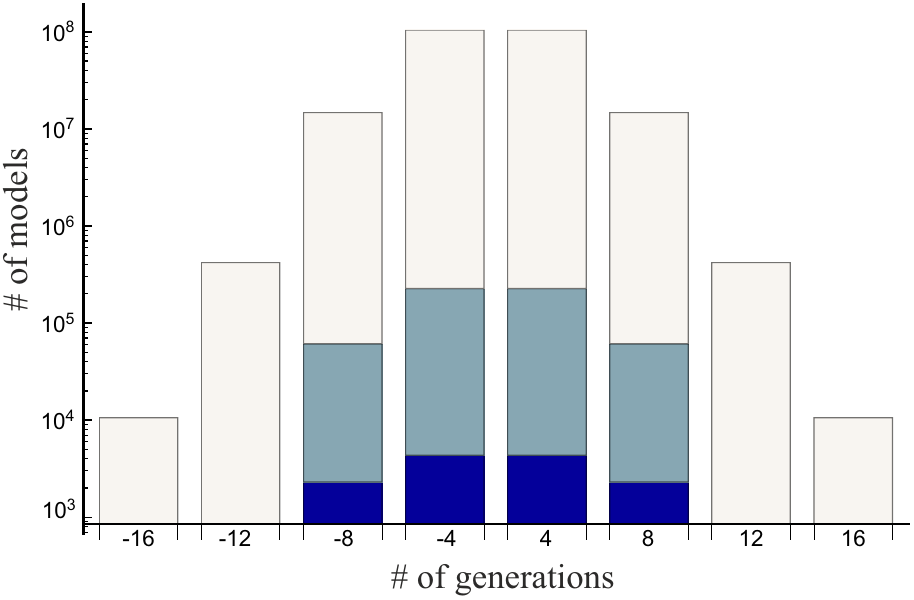}
\caption{Number of PS models versus chiral matter generations: Light shaded bars correspond to configurations satisfying criteria (a)-(e), whereas medium shading corresponds to models satisfying (a)-(g). Dark shaded bars represent the subset of models satisfying all previous criteria while simultaneously exhibiting a positive definite one-loop effective potential.}\label{oplot}
\end{figure}

The results of our computer search are outlined in Figure \ref{oplot}, where we plot the number of acceptable models versus the net number of fermion generations. With light shading we depict models that conform to requirements (a)-(e) as stated above, while medium shading corresponds to models that meet criteria (a)-(g). When comparing with the initial sample of $\sim 3.5\times10^{13}$ models, it should be noted that every model in Figure \ref{oplot} actually represents a family of $2^{11} = 2048$ models with identical characteristics, as far as the above criteria are concerned. In total, the light shaded region corresponds to $2.4\times10^8$ Pati--Salam models, that is one in approximately hundred models satisfies criteria (a) to (e). The  medium shaded plot region comprises $5.6\times10^5$ models, which means that only one in approximately 30,000 models meets all requirements, (a) to (g). In particular, this suggests that requirements (f),(g),
that is compatibility with the Scherk--Schwarz SUSY breaking mechanism and adherence to the super-no-scale constraints are hard to meet; they are satisfied only by two in one thousand models.

\FloatBarrier

	\FloatBarrier
	\begin{table}[h]\centering
		\def\sym#1{\ifmmode^{#1}\else\(^{#1}\)\fi}
		\begin{tabular}{l|c|c|c|c}
			\hline
			Class &$\abs{n_g}$& $n_H$& $n_h$& $\#$ of models \\
			\hline
			$A1$ &4 & 4 & 8 & 1536 \\
			\hline
			$A2$ &8 & 4 & 8 & 2048 \\
			\hline
		\end{tabular}
		\caption{\it Class A model synopsis.}
	\end{table}
	\FloatBarrier
	\FloatBarrier
	\begin{table}[h]\centering
		\def\sym#1{\ifmmode^{#1}\else\(^{#1}\)\fi}
		\begin{tabular}{l|c|c|c|c}
			\hline
			Class &$\abs{n_g}$& $n_H$& $n_h$& $\#$ of models \\
			\hline
			$B1$ & 4 & 4 & 4 & 3584 \\
			& 4 & 8 & 8 & 1792 \\
			& 8 & 4 & 8 & 4096 \\
			\hline
			$B2$ & 4 & 4 & 4 & 1792 \\
			\hline
		\end{tabular}
	\caption{\it Class B model synopsis.}
	\end{table}
	\FloatBarrier
	\FloatBarrier
	\begin{table}[h]\centering
		\def\sym#1{\ifmmode^{#1}\else\(^{#1}\)\fi}
			{\scriptsize
		\begin{tabular}{l|c|c|c|c}
			\hline
			Class &$\abs{n_g}$& $n_H$& $n_h$& $\#$ of models \\
			\hline
			$C1$ &  4 & 4 & 4 & 9984 \\
			&4 & 4 & 8 & 22784 \\
			&4 & 4 & 12 & 3584 \\
			&4 & 8 & 4 & 8192 \\
			&4 & 8 & 8 & 3072 \\
			&8 & 4 & 4 & 8192 \\
			&8 & 4 & 8 & 6144 \\
			&8 & 8 & 4 & 4096 \\
			&8 & 8 & 8 & 2048 \\ 
			\hline
			$C2$ &4 & 4 & 4 & 20480 \\
			&4 & 4 & 8 & 12544 \\
			&4 & 4 & 12 & 1792 \\
			&4 & 8 & 4 & 15104 \\
			&4 & 8 & 8 & 1792 \\
			&8 & 4 & 4 & 18432 \\
			&8 & 8 & 4 & 6144 \\
			\hline
			$C3$ &4 & 4 & 4 & 3584 \\
			&4 & 4 & 12 & 1792 \\
			&4 & 8 & 12 & 896 \\
			\hline
			$C4$ &4 & 4 & 8 & 1792 \\
			&8 & 4 & 8 & 2048 \\
			\hline
			$C5$ & 4 & 4 & 4 & 43264 \\
			&4 & 4 & 8 & 11776 \\
			&4 & 4 & 12 & 6144 \\
			&4 & 8 & 4 & 1792 \\
			&8 & 8 & 4 & 4096 \\
			&8 & 8 & 8 & 1024 \\
			\hline
			$C6$ &4 & 8 & 4 & 1792 \\
			&4 & 8 & 8 & 1792 \\
			\hline
			$C7$ &4 & 4 & 4 & 1792 \\
			&4 & 4 & 8 & 1792 \\
			&4 & 8 & 4 & 1792 \\
			&8 & 4 & 4 & 4096 \\
			\hline
			$C8$ & 4 & 4 & 4 & 12288 \\
			&4 & 4 & 8 & 9600 \\
			&4 & 4 & 12 & 6144 \\
			\hline
			$C9$ &4 & 4 & 4 & 6144 \\
			\hline
			$C10$ &4 & 4 & 4 & 1536 \\
			\hline
		\end{tabular}}
		\caption{\it Class C model synopsis.}
	\end{table}
	\FloatBarrier
	\FloatBarrier
	\begin{table}[h]\centering
		\def\sym#1{\ifmmode^{#1}\else\(^{#1}\)\fi}
		{\scriptsize
		\begin{tabular}{l|c|c|c|c}
			\hline
			Class &$\abs{n_g}$& $n_H$& $n_h$& $\#$ of models \\
			\hline
			$D1$ &8 & 8 & 8 & 3072 \\
			\hline
			$D2$ &4 & 4 & 8 & 8192 \\
			&8 & 8 & 8 & 3072 \\
			&8 & 16 & 8 & 1024 \\
			\hline
			$D3$ &4 & 4 & 4 & 61184 \\
			&4 & 4 & 8 & 34560 \\
			&4 & 4 & 12 & 3072 \\
			&4 & 4 & 16 & 1536 \\
			&4 & 8 & 4 & 5376 \\
			&8 & 4 & 4 & 2048 \\
			&8 & 8 & 4 & 2048 \\
			&8 & 8 & 8 & 14336 \\
			&8 & 16 & 8 & 512 \\
			\hline
			$D4$ &4 & 4 & 4 & 7168 \\
			&4 & 4 & 8 & 6656 \\
			&8 & 4 & 4 & 2048 \\
			&8 & 8 & 4 & 2048 \\
			\hline
			$D5$ &4 & 4 & 8 & 1152 \\
			\hline
			$D6$ & 4 & 4 & 8 & 1152 \\
			&8 & 8 & 8 & 768 \\
			\hline
			$D7$ &4 & 4 & 4 & 1792 \\
			\hline
			$D8$ &4 & 4 & 4 & 19712 \\
			&4 & 4 & 8 & 14336 \\
			&4 & 4 & 12 & 1792 \\
			&4 & 8 & 4 & 8704 \\
			&4 & 8 & 8 & 1792 \\
			&8 & 4 & 4 & 10240 \\
			&8 & 4 & 8 & 2048 \\
			&8 & 8 & 4 & 10240 \\
			&8 & 8 & 8 & 2048 \\
			\hline
			$D9$ &4 & 4 & 4 & 19520 \\
			&4 & 4 & 8 & 8960 \\
			&4 & 4 & 12 & 1344 \\
			&4 & 8 & 4 & 6528 \\
			&4 & 8 & 8 & 1344 \\
			\hline
			$D10$ &4 & 4 & 4 & 3584 \\
			&4 & 4 & 12 & 1792 \\
			&4 & 8 & 12 & 896 \\
			\hline
			$D11$ & 4 & 4 & 8 & 384 \\
			\hline
			$D12$ &4 & 4 & 8 & 224 \\
			&4 & 12 & 8 & 96 \\
			\hline
		\end{tabular}}
		\caption{\it Class D model synopsis.}
	\end{table}
	\FloatBarrier

An important aspect of our analysis concerns the study of the one-loop effective potential for all models satisfying the aforementioned criteria (a)-(g), and their subsequent classification according to the parameters $(V_{F}, \lambda_1)$ introduced in Section \ref{section3}. As explained there, the positivity property of the effective potential as a function of the Scherk-Schwarz modulus $T_2$, is dictated by both its value $V_F$ at the self-dual point, as well as by its asymptotic suppression at large volume, which is controlled by the Bose--Fermi degeneracy $\lambda_1$ at the first excited level (to be calculated at the generic point).

\begin{figure}[h!]
		\includegraphics[scale=.65]{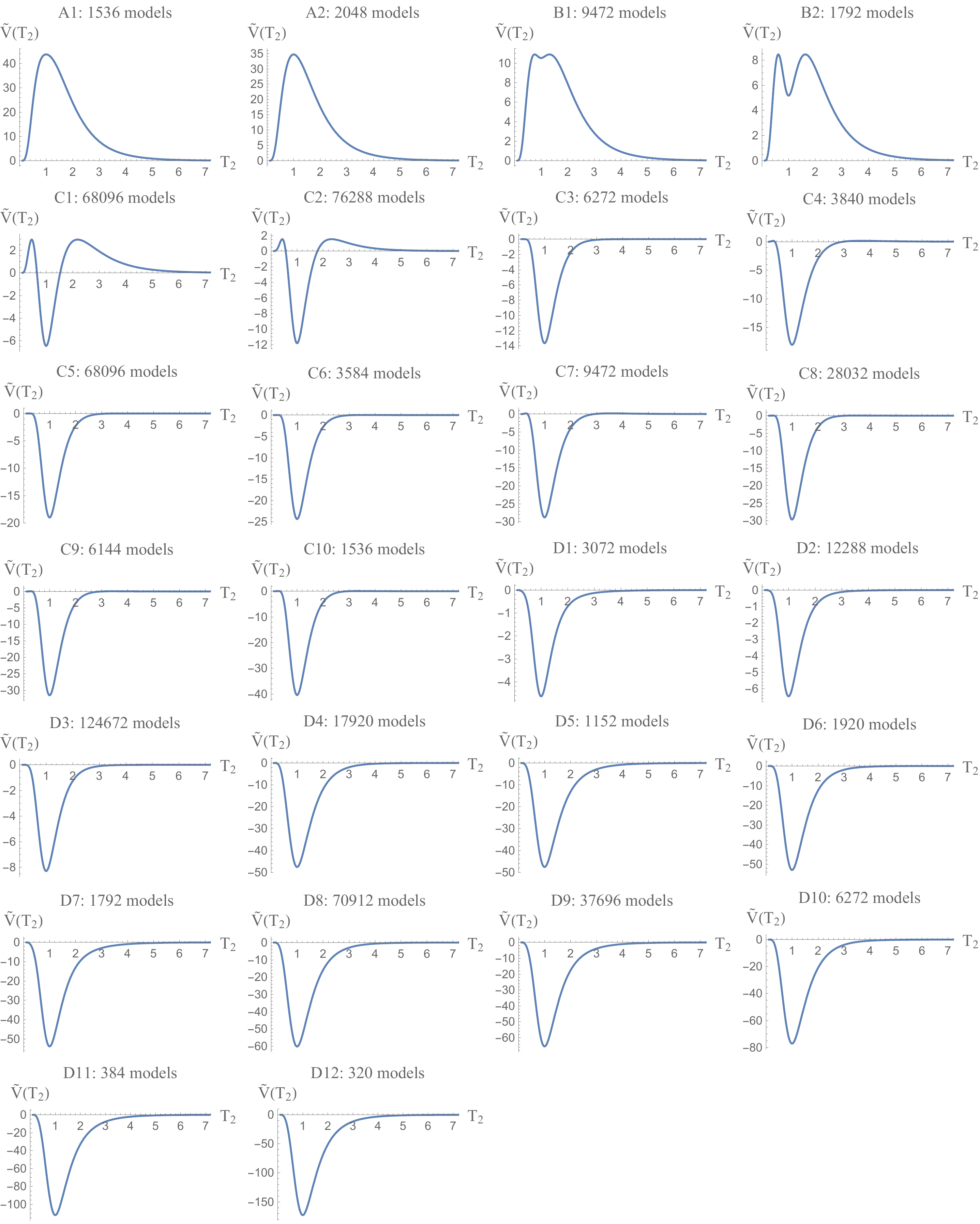}
		\caption{The rescaled effective potential $\tilde{V}(T_2)=2(2\pi)^4V(T_2)$ for each of the $26$ distinct subclasses of models. The number of models in each class is also displayed.}\label{fig:potentials1}
	\end{figure}

A careful study of the partition functions $Z(\tau,\bar\tau;t_I)$ allows us to group the models into 26 distinct subclasses. Indeed, each subclass corresponds to super no-scale PS models with identical partition functions, though not necessarily identical spectra. For each of these, we integrate the complex structure $\tau$ over the moduli space of the world-sheet torus, with each subclass giving rise to a distinct one-loop effective potential. As already mentioned in Section \ref{section3}, the non-BPS nature of the partition function, together with absolute convergence issues arising near self-dual points of the perturbative moduli space, effectively precludes an analytical treatment and instead necessitates a careful numerical evaluation of the modular integrals, which we carried out up to eighth order in the string mass level. Our results and classification of the potentials is shown in Figure \ref{fig:potentials1}. There, the models are organized into four broad classes, labelled $A$ through $D$, based on their shape and characteristics, which we now discuss in some detail.

The first class of models is denoted by $A$ and can be further split into two subclasses $A1, A2$. It contains models exhibiting a positive definite potential with a single global maximum at the fermionic point $T_2=1$, while the exponential decay at large volume is a consequence of the super no-scale property. A crucial feature is that this decay occurs while preserving the positivity of the potential, which is guaranteed by $V_F>0$ and $\lambda_1<0$. In addition to exhibiting the desired suppression, the potential dynamically pushes supersymmetry breaking to energy scales sufficiently lower than the string scale.

The second class, labelled $B$ in our classification, also splits into two subclasses $B1, B2$. While retaining their positivity and exponential decay $V_F>0, \lambda_1<0$, the corresponding potentials now exhibit a local minimum at the fermionic point, while the presence of a pair of global maxima at nearby ($T$-dual) values of the K\"ahler modulus is necessitated by the asymptotics and continuity. Although it might be tempting to consider the local minimum at $T_2=1$ as a metastable vaccum, one should be cautioned that regions of moduli space around self-dual points could turn unstable already at tree level. This is because fluctuations about such points typically give rise to tachyonic excitations as soon as the full parameter space of marginal  deformations is considered. Classes $A$ and $B$ contain a total of $1.5\times 10^4$ models and correspond to the dark shaded region in Figure \ref{oplot}.

The third class of models, denoted by $C$ is decomposed into 10 subclasses $C1,\ldots,C10$. Here, the potential at the fermionic point takes negative values $V_F<0$ and exhibits a global minimum, whereas $\lambda_1<0$  requires that the asymptotic decay at large volumes  still be attained from positive values. The positive definiteness of the potential is lost and the modulus is stabilized at the fermionic point. At face value, supersymmetry appears to be broken at the string scale, but the large negative cosmological constant will backreact against the tree-level geometry. Moreover, general fluctuations are again typically expected to trigger tachyonic Hagedorn-like instabilities that invalidate the perturbative analysis.

The final class of models, denoted by $D$, comprises of 12 subclasses $D1$ through $D12$ and corresponds to values $V_F<0$ and $\lambda_1>0$. These are negative definite potentials with a global minimum at the fermionic point, where an excess of bosonic degrees of freedom at the first massive level forces the potential to exponentially vanish from negative values in the asymptotic region $T_2\gg 1$. The modulus is again stabilized at the string scale and the problems of backreaction and tachyonic instabilities persist here as well, similarly to the previous discussion on class $C$.

\begin{figure}[h!]
		\includegraphics[scale=1]{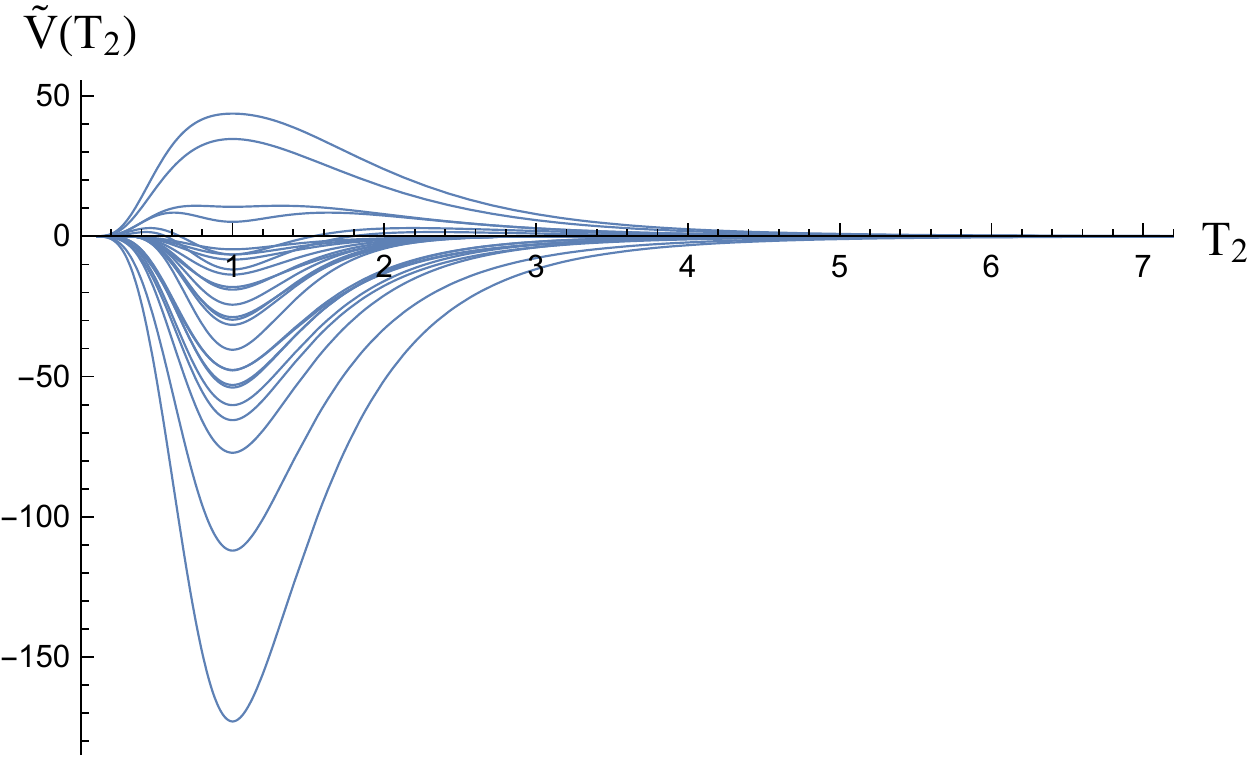}
		\caption{The rescaled effective potentials corresponding to the $26$ distinct subclasses of models.}\label{fig:potentials2}
	\end{figure}

The assembly of all potentials of classes $A,B,C$ and $D$ into a single plot  can be seen in Figure \ref{fig:potentials2}. Despite the fact that the super no-scale models in the distinct classes $A$ through $D$ are not continuously connected by marginal deformations, Figure \ref{fig:potentials2} suggests an underlying flow as $V_F$ (suitably rescaled) discontinuously varies from its maximal value $43.7351$ in subclass $A1$ to its miminal value $-172.913$ in $D12$. This  structure is a result of the effective modularity (T-duality) of the non-holomorphic, automorphic function arising from the theta lift \eqref{effective potential}, the super no-scale structure, as well as the asymptotic recovery of supersymmetry in the (infinite) decompactification limit, acting as a boundary condition for these curves.

	\FloatBarrier
	\FloatBarrier
	\FloatBarrier
	\FloatBarrier

At this stage, an interesting related question arises. Namely, we wish to ascertain how the structure of the one-loop potential is affected by the reduction of the observable gauge group. Indeed, the heterotic PS theories constructed in this work can be thought of as originating from a parent SO(10) theory with spontaneously broken $\mathcal N=1\to 0$ supersymmetry, in the class defined in \cite{Florakis:2016ani}.  In the formalism of the fermionic construction, the connection is easily obtained by simply removing the basis vector $\alpha$, along with its corresponding GGSO projections. This is equivalent to removing the $\mathbb Z_2$ sectors (and projector) labelled by the pair $(H',G')$ in the orbifold formulation \eqref{Orbifold PF}. Clearly, when going in the opposite direction, i.e. in reducing SO(10) models down to PS, we need to impose that the $\mathbb Z_2$ orbifold responsible for the breaking of the gauge group does not couple to any R-symmetry charge, otherwise supersymmetry would be explicitly broken in a trivial fashion. Taking these considerations into account, the PS and parent SO(10) models may then be studied separately along with the corresponding potentials. 

Although a systematic analysis along these lines does not present particular difficulty, here it will be sufficient to simply report the result. Perhaps unsurprisingly, we find that neither the super no-scale property, nor the positive definiteness of the one-loop potential are preserved under the $\mathbb Z_2$ reduction of the gauge group. This is the case, because both properties are highly sensitive to the total Bose-Fermi degeneracy of massless as well as massive degrees of freedom at the generic as well as at the fermionic point in the perturbative moduli space, and there is no particular structure nor mechanism that would \emph{a priori} cause the additional $\mathbb Z_2$ to respect any of those properties.

As a particular example, consider the PS model in subclass $A2$, defined in terms of the following GGSO matrix
\begin{equation}
	c[^{\beta_i}_{\beta_j}]=\left(
	\begin{array}{cccccccccc}
	+1 & +1 & -1 & +1 & +1 & +1 & +1 & +1 & +1 & +1 \\
	+1 & +1 & +1 & +1 & +1 & +1 & +1 & -1 & +1 & -1 \\
	-1 & +1 & +1 & -1 & -1 & +1 & +1 & +1 & -1 & +1 \\
	+1 & +1 & -1 & -1 & +1 & +1 & +1 & -1 & -1 & -1 \\
	+1 & +1 & -1 & +1 & -1 & +1 & -1 & -1 & -1 & +1 \\
	+1 & -1 & +1 & +1 & +1 & +1 & -1 & -1 & +1 & +1 \\
	+1 & -1 & +1 & +1 & -1 & -1 & +1 & +1 & +1 & +1 \\
	+1 & -1 & +1 & -1 & -1 & -1 & +1 & +1 & +1 & -1 \\
	+1 & +1 & -1 & -1 & -1 & +1 & +1 & +1 & +1 & +1 \\
	+1 & -1 & +1 & -1 & +1 & -1 & -1 & +1 & +1 & +1 \\
	\end{array}
	\right)
	\label{modela2}
\end{equation}
which, in addition to the super no-scale propertly, enjoys a positive definite one-loop potential. However, an analysis of its spectrum reveals that the SO(10) parent does not satisfy the super no-scale condition. Moreover, reintroducing the basis vector $\alpha$ and varying the $\alpha$-related GGSO coefficients gives rise to a variety of descendant PS models in which one may check explicitly that neither the super no-scale property, nor the shape of the potential are correlated to the those of its specific SO(10) parent. 

Nevertheless, a certain universality structure in the effective potentials emerges. Namely, the characteristic shapes of the one-loop potentials classified in \eqref{fig:potentials1} remain largely unchanged as one moves to the SO(10) level. Together with the arguments in Section \ref{section3}, which classify the shape of the potential essentially in terms of the two parameters $\lambda_1, V_F$, a comparison of the potentials at the SO(10) and Pati--Salam level supports the observation that the salient features of the shape of the one-loop potentials are to a large extent universal whenever supersymmetry is broken \`a la Scherk-Schwarz in heterotic theories, provided the super no-scale condition is met.

This observation reflects the fact that the string effective potential is globally sensitive to all degrees of freedom, including massive ones,  and a particular model  should probably not be excluded on the basis of its potential alone. Indeed, it is plausible that the addition or reshuffling of degrees of freedom and their masses by the introduction of additional orbifold factors, possibly involving the hidden sector, may eventually restore both the super no-scale property and the desired shape of the potential.


\section{Specific Models}
In this section we present three concrete examples of the non-supersymmetric Pati--Salam models discussed in Section \ref{section4}. For each model, we provide the detailed definition in the free fermionic formulation by means of its GGSO matrix, as well as the corresponding phase necessary for the orbifold realisation. With the exception of universal states, we present the massless spectra of fermionic and scalar matter, as well the partition function expansions at the fermionic point. These examples serve to better illustrate the structure of the models discussed. It should be noted that, by virtue of the phenomenological criteria of Section \ref{section2}, all these models exhibit absence of physical tachyons, complete chiral fermion generations, SM and PS gauge symmetry breaking Higgs scalars and do satisfy the super-no-scale condition. Nevertheless, their spectra cannot yet be considered as ``realistic", since, for instance, the number of fermion generations turns out to be a multiple of four. This is certainly not a surprise, and is mainly due to the fact that we group boundary conditions in terms of complexified world-sheet fermions. At present, this choice is primarily dictated for technical reasons, since the precise map between the free fermionic construction and the orbifold picture is then significantly simpler, and the space of models then allows for a comprehensive scan. The first two examples are typical models with positive potential that meet all phenomenological requirements of Section \ref{section2}. The third model also complies with all requirements, however, its one-loop potential belongs to the class $D3$, that is, it has a negative value minimum at the fermionic point $T_2=1$. We include it here because its spectrum exhibits an interesting property concerning exotic fractionally charged states.

As a first example, we present a model from class $A2$, referred below as Model 1 and defined by the GGSO matrix of Eq. \eqref{modela2}.
This model exhibits $8$ fermionic generations, from the sectors $S+b_1+T_3$, $S+b_2+T_1$, $S+b_1+T_2$ and $S+b_3$. The Standard Model breaking Higgs scalar can be accommodated in the bi-doublets arising in sectors $b_2+T_3+x$ and $b_3+T_1+T_2+x$, while the Pati--Salam gauge symmetry can be broken by giving VEVs to the neutral component of the scalars arising in sector $b_3+T_1$.
Furthermore, supersymmetry is spontaneously broken via the Scherk--Schwarz mechanism and the model enjoys massless Bose-Fermi degeneracy at generic points of the $T-U$ moduli space. The massless spectrum is presented in Tables \ref{TT6} and \ref{TT5}.
 
 \FloatBarrier
 \begin{table}[h]\centering
 	\def\sym#1{\ifmmode^{#1}\else\(^{#1}\)\fi}
 	\begin{tabular}{@{}l|l@{}}
 		\hline
 		Sector &$SU(4)\times SU(2)_L\times SU(2)_R\times U(1)^3\times SU(2)^4\times SO(8)$ representation(s)\\
 		\hline
 		$S+z_2$ & $({\bf 6}, {\bf 1}, {\bf 1}, 0, 0, 0, {\bf 1},{\bf 1},{\bf 1},{\bf 1},{\bf 8_c})\,$, $({\bf 1}, {\bf 1}, {\bf 1}, \pm1, 0, 0, {\bf 1},{\bf 1},{\bf 1},{\bf 1},{\bf 8_c})\,$ \\
 		& $({\bf 1}, {\bf 1}, {\bf 1}, 0, \pm1, 0, {\bf 1},{\bf 1},{\bf 1},{\bf 1},{\bf 8_c})\,$, $({\bf 1}, {\bf 1}, {\bf 1}, 0, 0, \pm1, {\bf 1},{\bf 1},{\bf 1},{\bf 1},{\bf 8_c})\,$ \\
 		\hline
 		$S+b_1+T_2$ & $4\times({\bf \bar{4}}, {\bf 1}, {\bf 2}, \frac{1}{2}, 0, 0, {\bf 1},{\bf 1},{\bf 1},{\bf 1},{\bf 1})\,$ \\
 		\hline
 		$S+b_1+T_3$ & $4\times({\bf 4}, {\bf 2}, {\bf 1}, \frac{1}{2}, 0, 0, {\bf 1},{\bf 1},{\bf 1},{\bf 1},{\bf 1})\,$ \\
 		\hline
 		$S+b_1+T_2+T_3+x$ & $8\times({\bf 1}, {\bf 1}, {\bf 1}, 0, \frac{1}{2}, \frac{1}{2}, {\bf 1},{\bf 1},{\bf 1},{\bf 1},{\bf 1})\,$, $8\times({\bf 1}, {\bf 1}, {\bf 1}, 0, -\frac{1}{2}, -\frac{1}{2}, {\bf 1},{\bf 1},{\bf 1},{\bf 1},{\bf 1})\,$ \\
 		\hline
 		$S+b_1+T_2+T_3+z_2+x$ & $4\times({\bf 1}, {\bf 1}, {\bf 1}, 0, \frac{1}{2}, -\frac{1}{2}, {\bf 1},{\bf 1},{\bf 1},{\bf 1},{\bf 8_s})\,$ \\
 		\hline
 		$S+b_1+T_3+\alpha$ & $4\times({\bf \bar{4}}, {\bf 1}, {\bf 1}, -\frac{1}{2}, 0, 0, {\bf 2},{\bf 1},{\bf 1},{\bf 1},{\bf 1})\,$ \\
 		\hline
 		$S+b_1+T_2+z_1+\alpha$ & $4\times({\bf 4}, {\bf 1}, {\bf 1}, -\frac{1}{2}, 0, 0, {\bf 1},{\bf 1},{\bf 2},{\bf 1},{\bf 1})\,$ \\
 		\hline
 		$S+b_2+T_1$ & $4\times({\bf 4}, {\bf 2}, {\bf 1}, 0, \frac{1}{2}, 0, {\bf 1},{\bf 1},{\bf 1},{\bf 1},{\bf 1})\,$ \\
 		\hline
 		$S+b_2+T_1+T_3+x$ & $4\times({\bf 1}, {\bf 2}, {\bf 2}, -\frac{1}{2}, 0, \frac{1}{2}, {\bf 1},{\bf 1},{\bf 1},{\bf 1},{\bf 1})\,$ \\
 		\hline
 		$S+b_2+z_1+x$ & $4\times({\bf 1}, {\bf 1}, {\bf 1}, \frac{1}{2}, 0, -\frac{1}{2}, {\bf 2},{\bf 1},{\bf 1},{\bf 2},{\bf 1})\,$ \\
 		\hline
 		$S+b_2+\alpha+x$ & $4\times({\bf 1}, {\bf 2}, {\bf 1}, -\frac{1}{2}, 0, -\frac{1}{2}, {\bf 1},{\bf 2},{\bf 1},{\bf 1},{\bf 1})\,$ \\
 		\hline
 		$S+b_2+T_1+T_3+z_1+\alpha+x$ & $4\times({\bf 1}, {\bf 2}, {\bf 1}, \frac{1}{2}, 0, -\frac{1}{2}, {\bf 1},{\bf 1},{\bf 2},{\bf 1},{\bf 1})\,$ \\
 		\hline
 		$S+b_3$ & $4\times({\bf \bar{4}}, {\bf 1}, {\bf 2}, 0, 0, -\frac{1}{2}, {\bf 1},{\bf 1},{\bf 1},{\bf 1},{\bf 1})\,$ \\
 		\hline
 		$S+b_3+T_2+x$ & $4\times({\bf 1}, {\bf 2}, {\bf 2}, -\frac{1}{2}, -\frac{1}{2}, 0, {\bf 1},{\bf 1},{\bf 1},{\bf 1},{\bf 1})\,$ \\
 		\hline
 		$S+b_3+T_1+z_1+x$ & $4\times({\bf 1}, {\bf 1}, {\bf 1}, \frac{1}{2}, \frac{1}{2}, 0, {\bf 1},{\bf 2},{\bf 2},{\bf 1},{\bf 1})\,$ \\
 		\hline
 		$S+b_3+T_2+\alpha+x$ & $4\times({\bf 1}, {\bf 1}, {\bf 2}, \frac{1}{2}, \frac{1}{2}, 0, {\bf 2},{\bf 1},{\bf 1},{\bf 1},{\bf 1})\,$ \\
 		\hline
 		$S+b_3+T_1+z_1+\alpha+x$ & $4\times({\bf 1}, {\bf 1}, {\bf 2}, -\frac{1}{2}, \frac{1}{2}, 0, {\bf 1},{\bf 1},{\bf 1},{\bf 2},{\bf 1})\,$ \\
 		\hline
 	\end{tabular}
 	\caption{\it Spectrum of massless fermionic matter (at the fermionic point) and quantum numbers under the gauge bundle for Model 1.\label{TT6}}
 \end{table}
 \FloatBarrier
 \FloatBarrier
 \begin{table}[h]\centering
 	\def\sym#1{\ifmmode^{#1}\else\(^{#1}\)\fi}
 	{
 		\begin{tabular}{@{}l|l@{}}
 			\hline
 			Sector &$SU(4)\times SU(2)_L\times SU(2)_R\times U(1)^3\times SU(2)^4\times SO(8)$ representation(s)\\
 			\hline
 			$0$ & $({\bf 6}, {\bf 1}, {\bf 1}, \pm1, 0, 0, {\bf 1},{\bf 1},{\bf 1},{\bf 1},{\bf 1})\,$, $({\bf 6}, {\bf 1}, {\bf 1}, 0, \pm1, 0, {\bf 1},{\bf 1},{\bf 1},{\bf 1},{\bf 1})\,$,\\
 			& $({\bf 6}, {\bf 1}, {\bf 1}, 0, 0, \pm1, {\bf 1},{\bf 1},{\bf 1},{\bf 1},{\bf 1})\,$, $({\bf 1}, {\bf 1}, {\bf 1}, \pm1, \pm1, 0, {\bf 1},{\bf 1},{\bf 1},{\bf 1},{\bf 1})\,$, \\
 			&$({\bf 1}, {\bf 1}, {\bf 1}, \pm1, 0, \pm1,{\bf 1},{\bf 1},{\bf 1},{\bf 1},{\bf 1})\,$, $({\bf 1}, {\bf 1}, {\bf 1}, 0, \pm1, \pm1, {\bf 1},{\bf 1},{\bf 1},{\bf 1},{\bf 1})\,$,\\
 			& $12\times({\bf 1}, {\bf 1}, {\bf 1}, 0, 0, 0, {\bf 1},{\bf 1},{\bf 1},{\bf 1},{\bf 1})\,$ \\
 			\hline
 			$\alpha$ & $4\times({\bf 1}, {\bf 2}, {\bf 1}, 0, 0, 0, {\bf 2},{\bf 1},{\bf 1},{\bf 1},{\bf 1})\,$\\
 			\hline
 			$z_1+\alpha$ & $4\times({\bf 1}, {\bf 1}, {\bf 2}, 0, 0, 0, {\bf 1},{\bf 1},{\bf 2},{\bf 1},{\bf 1})\,$ \\
 			\hline
 			$ b_1+T_2+T_3+x$ & $4\times({\bf 1}, {\bf 1}, {\bf 1}, 0, \frac{1}{2}, \frac{1}{2}, {\bf 1},{\bf 1},{\bf 1},{\bf 1},{\bf 8_v})\,$\\
 			\hline
 			$b_1+T_2+T_3+\alpha$ & $4\times({\bf 4}, {\bf 1}, {\bf 1}, \frac{1}{2}, 0, 0, {\bf 1},{\bf 2},{\bf 1},{\bf 1},{\bf 1})\,$\\
 			\hline
 			$b_1+T_2+T_3+z_1+\alpha$ & $4\times({\bf 4}, {\bf 1}, {\bf 1}, -\frac{1}{2}, 0, 0, {\bf 1},{\bf 1},{\bf 1},{\bf 2},{\bf 1})\,$\\
 			\hline
 			$b_2$ & $4\times({\bf 4}, {\bf 2}, {\bf 1}, 0, \frac{1}{2}, 0, {\bf 1},{\bf 1},{\bf 1},{\bf 1},{\bf 1})\,$\\
 			\hline
 			$b_2+T_3+x$ & $4\times({\bf 1}, {\bf 2}, {\bf 2}, -\frac{1}{2}, 0, \frac{1}{2}, {\bf 1},{\bf 1},{\bf 1},{\bf 1},{\bf 1})\,$\\
 			\hline
 			$b_2+T_1+z_1+x$ & $4\times({\bf 1}, {\bf 1}, {\bf 1}, \frac{1}{2}, 0, -\frac{1}{2}, {\bf 2},{\bf 1},{\bf 1},{\bf 2},{\bf 1})\,$\\
 			\hline
 			$b_2+T_1+\alpha+x$ & $4\times({\bf 1}, {\bf 2}, {\bf 1}, -\frac{1}{2}, 0 -\frac{1}{2}, {\bf 1},{\bf 2},{\bf 1},{\bf 1},{\bf 1})\,$\\
 			\hline
 			$b_2+T_3+z_1+\alpha+x$ & $4\times({\bf 1}, {\bf 2}, {\bf 1}, \frac{1}{2}, 0, -\frac{1}{2}, {\bf 1},{\bf 1},{\bf 2},{\bf 1},{\bf 1})\,$\\
 			\hline
 			$b_3+T_1$ & $4\times({\bf 4}, {\bf 1}, {\bf 2}, 0, 0, \frac{1}{2}, {\bf 1},{\bf 1},{\bf 1},{\bf 1},{\bf 1})\,$\\
 			\hline
 			$b_3+T_1+T_2+x$ & $4\times({\bf 1}, {\bf 2}, {\bf 2}, \frac{1}{2}, \frac{1}{2}, 0, {\bf 1},{\bf 1},{\bf 1},{\bf 1},{\bf 1})\,$\\
 			\hline
 			$b_3+z_1+x$ & $4\times({\bf 1}, {\bf 1}, {\bf 1}, \frac{1}{2}, \frac{1}{2}, 0, {\bf 1},{\bf 2},{\bf 2},{\bf 1},{\bf 1})\,$\\
 			\hline
 			$b_3+z_1+\alpha+x$ & $4\times({\bf 1}, {\bf 1}, {\bf 2}, -\frac{1}{2}, \frac{1}{2}, 0, {\bf 1},{\bf 1},{\bf 1},{\bf 2},{\bf 1})\,$\\
 			\hline
 			$b_3+T_1+T_2+\alpha+x$ & $4\times({\bf 1}, {\bf 1}, {\bf 2}, \frac{1}{2}, \frac{1}{2}, 0, {\bf 2},{\bf 1},{\bf 1},{\bf 1},{\bf 1})\,$\\
 			\hline
 	\end{tabular}}
 	\caption{\it Spectrum of massless scalar matter (at the fermionic point)  and quantum numbers under the  gauge bundle for Model 1.\label{TT5}}
 \end{table}
 \FloatBarrier

	The partition function at the fermionic point reads
\begin{equation}
\begin{aligned}
Z=&\frac{2q_i}{q_r}-\frac{16q_i}{\sqrt{q_r}}+\left(-24+128q_i+56q_i^2\right)+\left(1504+\frac{7424}{q_i}-1280q_i-416q_i^2\right)\sqrt{q_r} \\
&+\left(16384+\frac{17408}{q_i^2}+\frac{96384}{q_i}+3424q_i+1536q_i^2+792q_i^3\right)q_r+\mathcal{O}(q_r^{3/2}),
\end{aligned}
\end{equation}
where $q_i=e^{2\pi i\tau_1}$ and $q_r=e^{-2\pi\tau_2}$.
The partition function at generic points can be obtained from  Eq. \eqref{Orbifold PF} utilising the orbifold phase:
\begin{equation}
\begin{aligned}
\Phi=&ab+aG_1+bH_1+H_1G_1+kG+\ell H+\rho(G+G'+g_2+G_2)+\sigma(H+H'+h_2+H_2) \\
&+H(g_1+g_2+G_1)+G(h_1+h_2+H_1)+H'(g_2+G_1+G_3)+G'(h_2+H_1+H_3) +H'G' \\
&+h_1(G_2+G_3)+g_1(H_2+H_3)+h_2g_2+H_1G_2+H_2G_1+H_2G_2+H_2G_3+H_3G_2.
\end{aligned}
\end{equation}
This super-no-scale model meets all phenomenological requirements of Section \ref{section2} and bears a 
 positive semi-definite potential with a global maximum at the fermionic point and the desired exponential suppression at $T_2\gg1$.
\FloatBarrier
\begin{figure}[h!]
	\includegraphics[scale=1]{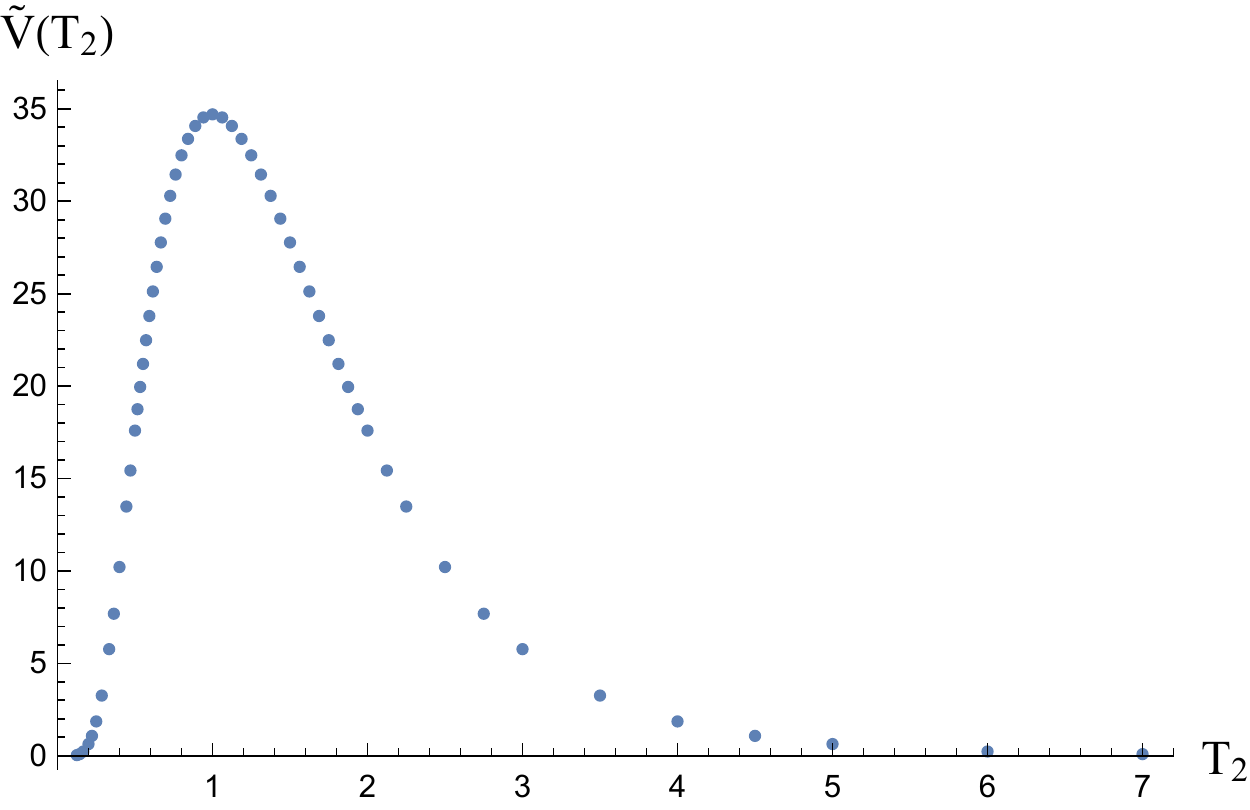}
	\caption{Numerical evaluation of the one-loop potential of Model 1 as a function of the $T_2$ modulus.}
\end{figure}
\FloatBarrier
	
The second example model of interest is selected from class $B2$, refered below as Model 2, corresponding to the following choice of GGSO phases	
	\begin{equation}
	c[^{\beta_i}_{\beta_j}]=\left(
	\begin{array}{cccccccccc}
	+1 & +1 & +1 & +1 & +1 & +1 & +1 & +1 & +1 & +1 \\
	+1 & +1 & +1 & +1 & -1 & +1 & +1 & +1 & -1 & +1 \\
	+1 & +1 & -1 & -1 & +1 & +1 & +1 & -1 & +1 & -1 \\
	+1 & +1 & -1 & -1 & +1 & +1 & +1 & +1 & -1 & +1 \\
	+1 & -1 & +1 & +1 & -1 & +1 & -1 & -1 & +1 & +1 \\
	+1 & -1 & +1 & +1 & +1 & +1 & +1 & -1 & -1 & -1 \\
	+1 & -1 & +1 & +1 & -1 & +1 & +1 & +1 & -1 & -1 \\
	+1 & +1 & -1 & +1 & -1 & -1 & +1 & +1 & +1 & -1 \\
	+1 & -1 & +1 & -1 & +1 & -1 & -1 & +1 & +1 & +1 \\
	+1 & +1 & -1 & +1 & +1 & +1 & +1 & +1 & +1 & +1 \\
	\end{array}
	\right)
	\end{equation}
Here, we have four fermion generations arising from sectors $S+b_1+T_2$ and $S+b_3$. The light (SM) and heavy (PS) Higgs bosons are obtained from the sectors $b_1+T_2+T_3+x$ and $b_3+T_1$, respectively. Additionally, generic points of the moduli space exhibit super no-scale structure, while supersymmetry breaking is compatible with the Scherk--Schwarz mechanism. The massless spectrum is presented in Tables \ref{TT8} and \ref{TT7}. 
	\FloatBarrier
	\begin{table}[h]\centering
		\def\sym#1{\ifmmode^{#1}\else\(^{#1}\)\fi}
		\begin{tabular}{@{}l|l@{}}
			\hline
			Sector &$SU(4)\times SU(2)_L\times SU(2)_R\times U(1)^3\times SU(2)^4\times SO(8)$ representation(s)\\
			\hline
			$S+\alpha$ & $4\times({\bf 1}, {\bf 2}, {\bf 1}, 0, 0, 0, {\bf 1},{\bf 2},{\bf 1},{\bf 1},{\bf 1})\,$\\
			\hline
			$S+z_2+\alpha$ & $({\bf 1}, {\bf 2}, {\bf 1}, 0, 0, 0, {\bf 1},{\bf 2},{\bf 1},{\bf 1},{\bf 8_s})\,$\\
			\hline
			$S+z_1+\alpha+x$ & $({\bf 4}, {\bf 1}, {\bf 1}, +\frac{1}{2}, -\frac{1}{2}, +\frac{1}{2}, {\bf 1},{\bf 1},{\bf 1},{\bf 2},{\bf 1})\,$, $({\bf 4}, {\bf 1}, {\bf 1}, +\frac{1}{2}, +\frac{1}{2}, -\frac{1}{2}, {\bf 1},{\bf 1},{\bf 1},{\bf 2},{\bf 1})\,$,\\ 
			& $({\bf 4}, {\bf 1}, {\bf 1}, -\frac{1}{2}, +\frac{1}{2}, +\frac{1}{2}, {\bf 1},{\bf 1},{\bf 1},{\bf 2},{\bf 1})\,$, $({\bf 4}, {\bf 1}, {\bf 1}, -\frac{1}{2}, -\frac{1}{2}, -\frac{1}{2}, {\bf 1},{\bf 1},{\bf 1},{\bf 2},{\bf 1})\,$,\\
			& $({\bf {\bar4}}, {\bf 1}, {\bf 1}, -\frac{1}{2},+\frac{1}{2}, -\frac{1}{2}, {\bf 1},{\bf 1},{\bf 1},{\bf 2},{\bf 1})\,$, $({\bf {\bar4}}, {\bf 1}, {\bf 1}, -\frac{1}{2}, -\frac{1}{2}, +\frac{1}{2}, {\bf 1},{\bf 1},{\bf 1},{\bf 2},{\bf 1})\,,$\\
			& $({\bf {\bar4}}, {\bf 1}, {\bf 1}, +\frac{1}{2}, -\frac{1}{2}, -\frac{1}{2}, {\bf 1},{\bf 1},{\bf 1},{\bf 2},{\bf 1})\,$, $({\bf {\bar4}}, {\bf 1}, {\bf 1}, +\frac{1}{2}, +\frac{1}{2}, +\frac{1}{2}, {\bf 1},{\bf 1},{\bf 1},{\bf 2},{\bf 1})\,$ \\
			\hline
			$S+b_1+T_2$ & $4\times({\bf 4}, {\bf 2}, {\bf 1},- \frac{1}{2}, 0, 0, {\bf 1},{\bf 1},{\bf 1},{\bf 1},{\bf 1})\,$\\
			\hline
			$S+b_1+x$ & $4\times({\bf 1}, {\bf 1}, {\bf 1}, 0, \frac{1}{2}, -\frac{1}{2}, {\bf 1},{\bf 1},{\bf 1},{\bf 1},{\bf 8_v})\,$\\
			\hline
			$S+b_1+T_2+x$ & $8\times({\bf 1}, {\bf 1}, {\bf 1}, 0, \frac{1}{2}, \frac{1}{2}, {\bf 1},{\bf 1},{\bf 1},{\bf 1},{\bf 1})\,$, $8\times({\bf 1}, {\bf 1}, {\bf 1}, 0, -\frac{1}{2}, -\frac{1}{2}, {\bf 1},{\bf 1},{\bf 1},{\bf 1},{\bf 1})\,$\\
			\hline
			$S+b_1+T_2+T_3+z_1+x$ & $4\times({\bf 1}, {\bf 1}, {\bf 1}, 0, \frac{1}{2}, -\frac{1}{2}, {\bf 1},{\bf 2},{\bf 2},{\bf 1},{\bf 1})\,$\\
			\hline
			$S+b_1+z_2+x$ & $4\times({\bf 1}, {\bf 1}, {\bf 1}, 0, -\frac{1}{2}, \frac{1}{2}, {\bf 1},{\bf 1},{\bf 1},{\bf 1},{\bf 8_c})\,$\\
			\hline
			$S+b_1+T_2+z_1+x$ & $4\times({\bf 1}, {\bf 1}, {\bf 1}, 0, -\frac{1}{2}, -\frac{1}{2}, {\bf 1},{\bf 2},{\bf 1},{\bf 2},{\bf 1})\,$\\
			\hline
			$S+b_1+T_2+\alpha+x$ & $4\times({\bf 1}, {\bf 2}, {\bf 1}, 0, \frac{1}{2}, -\frac{1}{2}, {\bf 2},{\bf 1},{\bf 1},{\bf 1},{\bf 1})\,$\\
			\hline
			$S+b_1+T_2+T_3+\alpha+x$ & $4\times({\bf 1}, {\bf 1}, {\bf 2}, 0, -\frac{1}{2}, \frac{1}{2}, {\bf 1},{\bf 2},{\bf 1},{\bf 1},{\bf 1})\,$\\
			\hline
			$S+b_2+T_3+x$ & $4\times({\bf 1}, {\bf 1}, {\bf 1}, -\frac{1}{2}, 0, -\frac{1}{2}, {\bf 1},{\bf 1},{\bf 1},{\bf 1},{\bf 8_v})\,$\\
			\hline
			$S+b_3$ & $4\times({\bf {\bar 4}}, {\bf 1}, {\bf 2}, 0, 0, -\frac{1}{2}, {\bf 1},{\bf 1},{\bf 1},{\bf 1},{\bf 1})\,$\\
			\hline
			$S+b_3+z_1+x$ & $4\times({\bf 1}, {\bf 1}, {\bf 1}, -\frac{1}{2}, -\frac{1}{2}, 0, {\bf 2},{\bf 1},{\bf 2},{\bf 1},{\bf 1})\,$\\
			\hline
			$S+b_3+T_1+z_1+x$ & $4\times({\bf 1}, {\bf 1}, {\bf 1}, -\frac{1}{2}, \frac{1}{2}, 0, {\bf 1},{\bf 2},{\bf 2},{\bf 1},{\bf 1})\,$\\
			\hline
			$S+b_3+z_1+\alpha+x$ & $4\times({\bf 1}, {\bf 2}, {\bf 1}, \frac{1}{2}, -\frac{1}{2}, 0, {\bf 1},{\bf 1},{\bf 2},{\bf 1},{\bf 1})\,$\\
			\hline
			$S+b_3+T_1+z_1+\alpha+x$ & $4\times({\bf 1}, {\bf 1}, {\bf 2}, -\frac{1}{2}, \frac{1}{2}, 0, {\bf 1},{\bf 1},{\bf 1},{\bf 2},{\bf 1})\,$\\
			\hline
		\end{tabular}
		\caption{\it Spectrum of massless fermionic matter and quantum numbers under the gauge bundle for Model 2.\label{TT8}}
	\end{table}
	\FloatBarrier
	\FloatBarrier
	\begin{table}[h]\centering
		\def\sym#1{\ifmmode^{#1}\else\(^{#1}\)\fi}
		\begin{tabular}{@{}l|l@{}}
			\hline
			Sector &$SU(4)\times SU(2)_L\times SU(2)_R\times U(1)^3\times SU(2)^4\times SO(8)$ representation(s)\\
			\hline
			$0$ & $({\bf 6}, {\bf 1}, {\bf 1}, \pm1, 0, 0, {\bf 1},{\bf 1},{\bf 1},{\bf 1},{\bf 1})\,$, $({\bf 6}, {\bf 1}, {\bf 1}, 0, \pm1, 0, {\bf 1},{\bf 1},{\bf 1},{\bf 1},{\bf 1})\,$,\\
			& $({\bf 6}, {\bf 1}, {\bf 1}, 0, 0, \pm1, {\bf 1},{\bf 1},{\bf 1},{\bf 1},{\bf 1})\,$, $({\bf 1}, {\bf 1}, {\bf 1}, \pm1, \pm1, 0, {\bf 1},{\bf 1},{\bf 1},{\bf 1},{\bf 1})\,$, \\
			&$({\bf 1}, {\bf 1}, {\bf 1}, \pm1, 0, \pm1,{\bf 1},{\bf 1},{\bf 1},{\bf 1},{\bf 1})\,$, $({\bf 1}, {\bf 1}, {\bf 1}, 0, \pm1, \pm1, {\bf 1},{\bf 1},{\bf 1},{\bf 1},{\bf 1})\,$,\\
			& $12\times({\bf 1}, {\bf 1}, {\bf 1}, 0, 0, 0, {\bf 1},{\bf 1},{\bf 1},{\bf 1},{\bf 1})\,$ \\
			\hline
			$z_1$ & $4\times({\bf 1}, {\bf 1}, {\bf 1}, 0, 0, 0, {\bf 1},{\bf 2},{\bf 1},{\bf 2},{\bf 1})\,$\\
			\hline
			$z_2$ & $4\times({\bf 1}, {\bf 1}, {\bf 1}, 0, 0, 0, {\bf 1},{\bf 1},{\bf 1},{\bf 1},{\bf 8_s})\,$\\
			\hline
			$\alpha$ & $({\bf 1}, {\bf 2}, {\bf 1}, 0, 0, \pm1, {\bf 1},{\bf 2},{\bf 1},{\bf 1},{\bf 1})\,$, $({\bf 1}, {\bf 2}, {\bf 1}, 0, \pm1, 0, {\bf 1},{\bf 2},{\bf 1},{\bf 1},{\bf 1})\,$\\
			\hline
			$z_1+\alpha$ & $4\times({\bf 1}, {\bf 1}, {\bf 2}, 0, 0, 0, {\bf 1},{\bf 1},{\bf 2},{\bf 1},{\bf 1})\,$\\
			\hline
			$b_1+T_2+x$ & $4\times({\bf 1}, {\bf 1}, {\bf 1}, 0, -\frac{1}{2}, +\frac{1}{2}, {\bf 2},{\bf 2},{\bf 1},{\bf 1},{\bf 1})\,$\\
			\hline
			$b_1+T_2+T_3+x$ & $4\times({\bf 1}, {\bf 2}, {\bf 2}, 0, \frac{1}{2}, -\frac{1}{2}, {\bf 1},{\bf 1},{\bf 1},{\bf 1},{\bf 1})\,$\\
			\hline
			$b_1+T_2+\alpha$ & $4\times({\bf 4}, {\bf 1}, {\bf 1}, -\frac{1}{2}, 0, 0, {\bf 1},{\bf 2},{\bf 1},{\bf 1},{\bf 1})\,$\\
			\hline
			$b_1+T_2+z_1+\alpha+x$ & $4\times({\bf 1}, {\bf 2}, {\bf 1}, 0, +\frac{1}{2}, +\frac{1}{2}, {\bf 1},{\bf 1},{\bf 1},{\bf 2},{\bf 1})\,$\\
			\hline
			$b_1+T_2+T_3+z_1+\alpha+x$ & $4\times({\bf 1}, {\bf 2}, {\bf 1}, 0, -\frac{1}{2}, +\frac{1}{2}, {\bf 1},{\bf 1},{\bf 2},{\bf 1},{\bf 1})\,$\\
			\hline
			$b_2+T_1+T_3+x$ &$4\times({\bf 1}, {\bf 1}, {\bf 1}, +\frac{1}{2}, 0, +\frac{1}{2}, {\bf 1},{\bf 1},{\bf 1},{\bf 1},{\bf 8_v})\,$\\
			\hline
			$b_3+T_1$ & $4\times({\bf 4}, {\bf 1}, {\bf 2}, 0, 0, +\frac{1}{2}, {\bf 1},{\bf 1},{\bf 1},{\bf 1},{\bf 1})\,$\\
			\hline
			$b_3+z_1+x$ & $4\times({\bf 1}, {\bf 1}, {\bf 1}, \frac{1}{2}, -\frac{1}{2}, 0, {\bf 1},{\bf 2},{\bf 2},{\bf 1},{\bf 1})\,$\\
			\hline
			$b_3+T_1+z_1+x$ & $4\times({\bf 1}, {\bf 1}, {\bf 1}, \frac{1}{2}, \frac{1}{2}, 0, {\bf 2},{\bf 1},{\bf 2},{\bf 1},{\bf 1})\,$\\
			\hline
			$b_3+z_1+\alpha+x$ & $4\times({\bf 1}, {\bf 1}, {\bf 2}, \frac{1}{2}, -\frac{1}{2}, 0, {\bf 1},{\bf 1},{\bf 1},{\bf 2},{\bf 1})\,$\\
			\hline
			$b_3+T_1+z_1+\alpha+x$ & $4\times({\bf 1}, {\bf 2}, {\bf 1}, -\frac{1}{2}, \frac{1}{2}, 0, {\bf 1},{\bf 1},{\bf 2},{\bf 1},{\bf 1})\,$\\
			\hline
		\end{tabular}
		\caption{\it Spectrum of massless scalar matter (at the fermionic point) and quantum numbers under the  gauge bundle for Model 2. \label{TT7}}
	\end{table}
	\FloatBarrier
	The partition function at the fermionic point reads:
	\begin{equation}
	\begin{aligned}
	Z=&\frac{2q_i}{q_r}-\frac{16q_i}{\sqrt{q_r}}+\left(40+\frac{48}{q_i}+144q_i+56q_i^2\right)+\left(1248+\frac{6528}{q_i}-1152q_i-416q_i^2\right)\sqrt{q_r} \\
	&+\left(8192+\frac{18816}{q_i^2}+\frac{105600}{q_i}+4960q_i+2688q_i^2+792q_i^3\right)q_r+\mathcal{O}(q_r^{3/2})
	\end{aligned}
	\end{equation}
	and the phase entering the orbifold partition function Eq. \eqref{Orbifold PF}, reads
	\begin{equation}
	\begin{aligned}
	\Phi=&ab+aG_1+bH_1+H_1G_1+k(g_1+G_1+G_3)+\ell(h_1+H_1+H_3)+k\ell+\rho(G'+G_1+G_3) \\
		&+\sigma(H'+H_1+H_3)+H(G'+g_1+g_2+G_1+G_2)+G(H'+h_1+h_2+H_1+H_2)\\
			&+H'G_2+G'H_2+H'G'+h_1G_1+g_1H_1+h_1g_1 +h_2(G_1+G_3)+g_2(H_1+H_3)+H_1G_2+H_2G_1.
	\end{aligned}
	\end{equation}
	The one-loop effective potential remains positive semi-definite as we vary $T_2$, however in this case instead of a global maximum, the fermionic point corresponds to a local minimum, as shown in the following figure.
	\FloatBarrier
	\begin{figure}[h!]
		\includegraphics[scale=1]{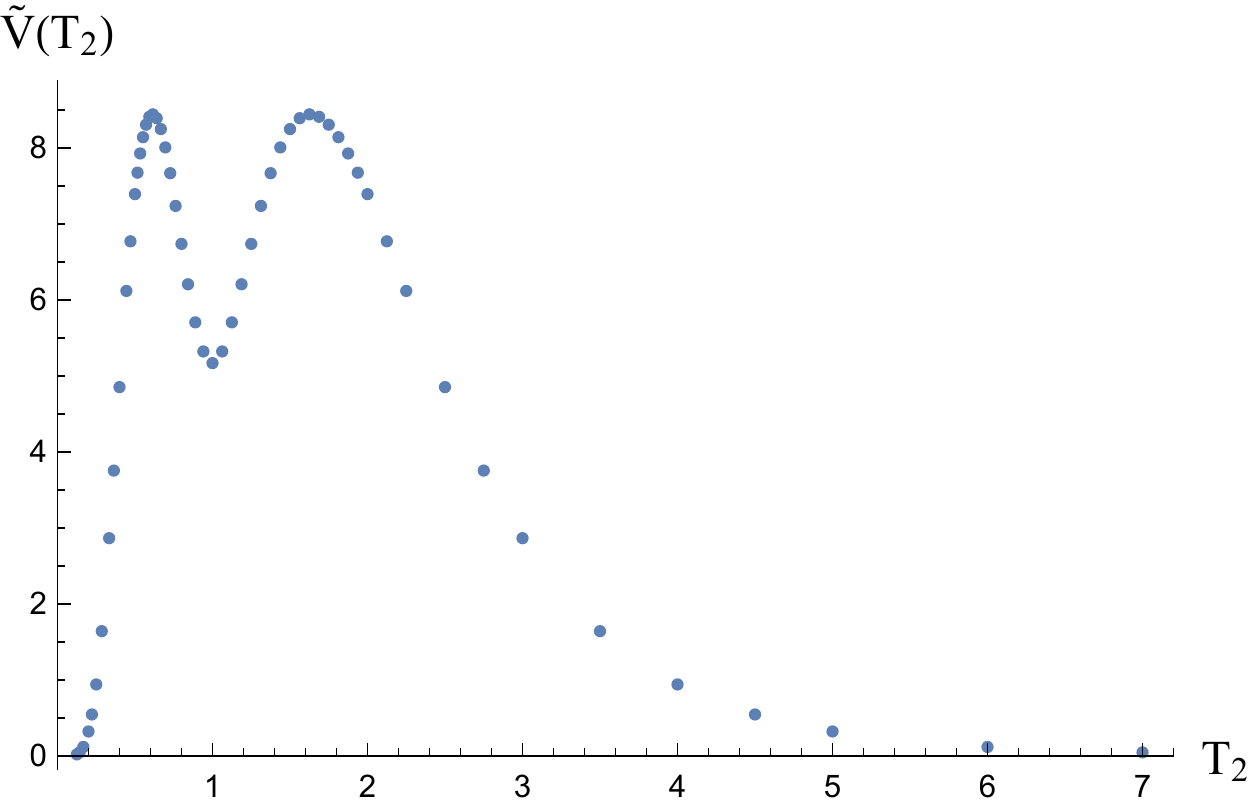}
		\caption{Numerical evaluation of the one-loop potential of Model 2.}
	\end{figure}
	\FloatBarrier

The spectra of both Model 1 and Model 2 contain fractionally charged exotic states transforming as $(\bf{4}+\bf{\bar{4}},\bf {1},\bf{1})$ or $(\bf{1},\bf {2},\bf{1})$/$(\bf{1},\bf{1},\bf{2})$ under the observable Pati--Salam gauge group $SU(4)\times{SU(2)}_L\times{SU(2)}_R$ along the lines discussed in Section  \ref{section2}. Of course, these states are vector-like, following the phenomenological requirements of Section \ref{section4}, and are expected to become super-heavy. Whereas it has been demonstrated that they can be eliminated completely from the massless spectrum in the case of supersymmetric Pati--Salam vacua defined in terms of non-complexifiable fermionic coordinates  \cite{Assel:2009xa}, it remains an open question whether this scenario can be also realised in the case of non-supersymmetric theories.

An interesting scenario emerges upon consideration of certain models exhibiting special gauge group enhancement. Namely, there exist gauge enhanced models where  all fractionally charged exotic states transform non-trivially under a potentially ``strong" hidden sector gauge group factor. By ``strong" we mean here a gauge group factor with gauge coupling running faster than $SU(3)$ that can a priori lead to exotic state confinement. A typical model of this kind, encountered in class $D3$, that we refer to as Model 3, is defined by the following GGSO matrix:
	\begin{equation}\label{enhanced model GGSO}
	c\begin{bsmallmatrix}
	\beta_i \\ \beta_j
	\end{bsmallmatrix}=
	\left(
	\begin{array}{cccccccccc}
	+1 & +1 & -1 & -1 & -1 & +1 & +1 & +1 & +1 & -1 \\
	+1 & +1 & +1 & +1 & -1 & +1 & +1 & -1 & -1 & +1 \\
	-1 & +1 & +1 & -1 & +1 & +1 & +1 & +1 & +1 & -1 \\
	-1 & +1 & -1 & +1 & +1 & +1 & +1 & +1 & +1 & -1 \\
	-1 & -1 & +1 & +1 & +1 & +1 & +1 & +1 & +1 & +1 \\
	+1 & -1 & +1 & +1 & +1 & +1 & +1 & +1 & +1 & -1 \\
	+1 & -1 & +1 & +1 & +1 & +1 & +1 & +1 & +1 & +1 \\
	+1 & -1 & +1 & +1 & +1 & +1 & +1 & +1 & +1 & -1 \\
	+1 & -1 & +1 & +1 & +1 & +1 & +1 & +1 & +1 & -1 \\
	-1 & +1 & -1 & -1 & +1 & +1 & -1 & +1 & -1 & -1 \\
	\end{array}
	\right)
	\end{equation}
	The resulting theory exhibits spontaneously broken supersymmetry \`a la Scherk--Schwarz as well as massless Bose-Fermi degeneracy at generic points of the $T-U$ moduli space, while its spectrum satisfies our imposed phenomenological constraints. However, it also exhibits gauge symmetry enhancement due to vector bosons arising from  $z_1$ and $z_1+z_2$ sectors. The full gauge symmetry of this model is 
	\begin{align}
	G=\{SU(4)\times SU(2)_L\times SU(2)_R\}_{\rm observable}\times U(1)^2\times SO(6)\times SO(12)\,.
	\end{align}
	The fermionic point partition function is given by
	\begin{equation}
	\begin{aligned}
	Z=&\frac{2q_i}{q_r}-\frac{16q_i}{\sqrt{q_r}}+\left(8-32q_i+56q_i^2\right)+\left(4064+\frac{4096}{q_i}+512q_i-416q_i^2\right)\sqrt{q_r} \\
	&+\left(\frac{16384}{q_i^2}+\frac{122880}{q_i}-8480q_i-768q_i^2+792q_i^3\right)q_r+\mathcal{O}(q_r^{3/2})\,,
	\end{aligned}
	\end{equation}
	while the corresponding orbifold phase reads
	\begin{equation}
	\begin{aligned}\label{enhanced model phase}
	\Phi=&ab+aG_1+bH_1+H_1G_1+k(G+g_1+g_2)+\ell(H+h_1+h_2)+\rho G'+\sigma H' \\
	&+H(G'+g_1+g_2)+G(H'+h_1+h_2)+HG+H'(g_1+G_1)+G'(h_1+H_1)+H'G'\\
	&+h_1g_1+h_2g_2+H_1G_2+H_2G_1\,.
	\end{aligned}
	\end{equation}
	The full massless spectrum of Model 3 is presented in Tables \ref{TT10} and \ref{TT9}.
	\FloatBarrier
	\begin{table}[h]\centering
		\def\sym#1{\ifmmode^{#1}\else\(^{#1}\)\fi}
		{\small
			\begin{tabular}{@{}l|l@{}}
				\hline
				Sector(s) &$SU(4)\times SU(2)_L\times SU(2)_R\times U(1)^2\times SO(6)\times SO(12)$ representation(s)\\
				\hline
				$S+\alpha (+z_1+z_2)$ & $({\bf 1}, {\bf 2}, {\bf 1}, 0, 0,{\bf 1},{\bf 32})\,$\\
				\hline 
				$S+\alpha+z_1 (+z_1+z_2)$  & $({\bf 1}, {\bf 1}, {\bf 2}, 0, 0,{\bf 1},{\bf 32})\,$\\
				\hline
				$S+x+\alpha (+z_1)$ & $({\bf 4}, {\bf 1}, {\bf 1}, \pm\frac{1}{2}, \pm\frac{1}{2},{\bf \bar{4}},{\bf 1})\,$, $({\bf \bar{4}}, {\bf 1}, {\bf 1}, \pm\frac{1}{2}, \pm\frac{1}{2},{\bf 4},{\bf 1})\,$\\
				\hline
				$S+b_1+T_2$ & $4\times({\bf \bar{4}}, {\bf 1}, {\bf 2}, -\frac{1}{2}, 0,{\bf 1},{\bf 1})\,$\\\hline
				$S+b_1+T_2+T_3$ & $4\times({\bf \bar{4}}, {\bf 1}, {\bf 2}, -\frac{1}{2}, 0,{\bf 1},{\bf 1})\,$\\\hline
				$S+b_1+x+\alpha (+z_1)$ & $4\times({\bf 1}, {\bf 1}, {\bf 2}, 0, -\frac{1}{2},{\bf \bar{4}},{\bf 1})\,$\\
				\hline
				$S+b_1+T_3+x+\alpha (+z_1)$ & $4\times({\bf 1}, {\bf 1}, {\bf 2}, 0, -\frac{1}{2},{\bf \bar{4}},{\bf 1})\,$\\
				\hline
				$S+b_2+T_1$ & $4\times({\bf 4}, {\bf 2}, {\bf 1}, 0, -\frac{1}{2},{\bf 1},{\bf 1})\,$\\\hline
				$S+b_2+T_1+T_3$ & $4\times({\bf 4}, {\bf 2}, {\bf 1}, 0, -\frac{1}{2},{\bf 1},{\bf 1})\,$\\\hline
				$S+b_2+x+\alpha (+z_1)$ & $4\times({\bf 1}, {\bf 2}, {\bf 1}, -\frac{1}{2}, 0,{\bf 4},{\bf 1})\,$\\
				\hline
				$S+b_2+T_3+x+\alpha (+z_1)$ & $4\times({\bf 1}, {\bf 2}, {\bf 1}, -\frac{1}{2}, 0,{\bf 4},{\bf 1})\,$\\
				\hline
				$S+b_3+T_1+x$ & $4\times({\bf 1}, {\bf 2}, {\bf 2}, \frac{1}{2}, \frac{1}{2},{\bf 1},{\bf 1})\,$\\\hline
				$S+b_3+T_2+x$ & $4\times({\bf 1}, {\bf 2}, {\bf 2}, \frac{1}{2}, \frac{1}{2},{\bf 1},{\bf 1})\,$\\\hline
				$S+b_3+x (+z_1)$ & $4\times({\bf 6}, {\bf 1}, {\bf 1}, \frac{1}{2}, \frac{1}{2},{\bf 1},{\bf 1})\,$, $4\times({\bf 1}, {\bf 1}, {\bf 1}, -\frac{1}{2}, -\frac{1}{2},{\bf 6},{\bf 1})\,$, $4\times({\bf 1}, {\bf 1}, {\bf 1}, -\frac{1}{2}, \frac{1}{2},{\bf 1},{\bf 1})\,$,\\
				& $4\times({\bf 1}, {\bf 1}, {\bf 1}, \frac{1}{2}, -\frac{1}{2},{\bf 1},{\bf 1})\,$\\\hline
				$S+b_3+T_1+T_2+x (+z_1)$ & $4\times({\bf 6}, {\bf 1}, {\bf 1}, -\frac{1}{2}, -\frac{1}{2},{\bf 1},{\bf 1})\,$, $4\times({\bf 1}, {\bf 1}, {\bf 1}, \frac{1}{2}, \frac{1}{2},{\bf 6},{\bf 1})\,$, $4\times({\bf 1}, {\bf 1}, {\bf 1}, -\frac{1}{2}, \frac{1}{2},{\bf 1},{\bf 1})\,$,\\
				& $4\times({\bf 1}, {\bf 1}, {\bf 1}, \frac{1}{2}, -\frac{1}{2},{\bf 1},{\bf 1})\,$\\\hline
				$S+b_3+T_1+\alpha (+z_1)$ & $4\times({\bf 4}, {\bf 1}, {\bf 1}, 0, 0,{\bf \bar{4}},{\bf 1})\,$\\
				\hline
				$S+b_3+T_2+\alpha (+z_1)$ & $4\times({\bf \bar{4}}, {\bf 1}, {\bf 1}, 0, 0,{\bf 4},{\bf 1})\,$\\
				\hline
				$S+b_3+T_1+x+z_1 (+z_1+z_2)$ & $4\times({\bf 1}, {\bf 1}, {\bf 1}, -\frac{1}{2}, -\frac{1}{2},{\bf 1},{\bf 12})\,$\\
				\hline
				$S+b_3+T_2+x+z_1 (+z_1+z_2)$ & $4\times({\bf 1}, {\bf 1}, {\bf 1}, -\frac{1}{2}, -\frac{1}{2},{\bf 1},{\bf 12})\,$\\
				\hline
		\end{tabular}}
		\caption{\it Spectrum of massless fermionic matter and quantum numbers under the gauge bundle for Model 3.  Multiplets of the hidden $SO(6)$ and $SO(12)$ gauge groups are obtained from sector pairs of the form  $\xi (+z_1) = \{\xi, \xi+z_1\}$ and $\xi(+z_1+z_2)=\{\xi, \xi+z_1+z_2\}$ respectively. \label{TT10}}
	\end{table}
	\FloatBarrier
	\FloatBarrier
	\begin{table}[h]\centering
		\def\sym#1{\ifmmode^{#1}\else\(^{#1}\)\fi}
		{\small
			\begin{tabular}{@{}l|l@{}}
				\hline
				Sector(s) &$SU(4)\times SU(2)_L\times SU(2)_R\times U(1)^2\times SO(6)\times SO(12)$ representation(s)\\
				\hline
				$0$, $z_1, z_2$ & $({\bf 6}, {\bf 1}, {\bf 1}, 0, 0,{\bf 6},{\bf 1})\,$, $({\bf 6}, {\bf 1}, {\bf 1}, \pm1, 0,{\bf 1},{\bf 1})\,$, $({\bf 6}, {\bf 1}, {\bf 1}, 0, \pm1,{\bf 1},{\bf 1})\,$, \\
				& $({\bf 1}, {\bf 1}, {\bf 1}, \pm 1, 0,{\bf 6},{\bf 1})\,$, $({\bf 1}, {\bf 1}, {\bf 1}, 0, \pm1,{\bf 6},{\bf 1})\,$, $({\bf 1}, {\bf 1}, {\bf 1}, \pm1, \pm1,{\bf 1},{\bf 1})\,$,\\
				& $12\times({\bf 1}, {\bf 1}, {\bf 1}, 0, 0,{\bf 1},{\bf 1})\,$,
				$({\bf 1}, {\bf 2}, {\bf 2}, 0, 0,{\bf 1},{\bf 12})\,$ \\\hline
				$b_1$ & $4\times({\bf 4}, {\bf 1}, {\bf 2}, \frac{1}{2}, 0,{\bf 1},{\bf 1})\,$\\
				\hline
				$b_1+T_3$ & $4\times({\bf 4}, {\bf 1}, {\bf 2}, \frac{1}{2}, 0,{\bf 1},{\bf 1})\,$\\
				\hline
				$b_1+T_2+x+\alpha (+z_1)$  & $4\times({\bf 1}, {\bf 1}, {\bf 2}, 0, \frac{1}{2},{\bf 4},{\bf 1})\,$\\
				\hline
				$b_1+T_2+T_3+\alpha (+z_1)$  & $4\times({\bf 1}, {\bf 1}, {\bf 2}, 0, \frac{1}{2},{\bf 4},{\bf 1})\,$\\
				\hline
				$b_2$ & $4\times({\bf 4}, {\bf 2}, {\bf 1}, 0, -\frac{1}{2},{\bf 1},{\bf 1})\,$\\
				\hline
				$b_2+T_3$ & $4\times({\bf 4}, {\bf 2}, {\bf 1}, 0, -\frac{1}{2},{\bf 1},{\bf 1})\,$\\
				\hline
				$b_2+T_1+x+\alpha (+z_1)$ & $4\times({\bf 1}, {\bf 2}, {\bf 1}, -\frac{1}{2}, 0,{\bf 4},{\bf 1})\,$\\
				\hline
				$b_2+T_1+T_3+x+\alpha (+z_1)$ & $4\times({\bf 1}, {\bf 2}, {\bf 1}, -\frac{1}{2}, 0,{\bf 4},{\bf 1})\,$\\
				\hline
				$b_3+x$ & $4\times({\bf 1}, {\bf 2}, {\bf 2}, \frac{1}{2}, \frac{1}{2},{\bf 1},{\bf 1})\,$\\
				\hline
				$b_3+T_1+x (+z_1)$ & $4\times({\bf 6}, {\bf 1}, {\bf 1}, \frac{1}{2}, \frac{1}{2},{\bf 1},{\bf 1})\,$, $4\times({\bf 1}, {\bf 1}, {\bf 1}, \frac{1}{2}, \frac{1}{2},{\bf 6},{\bf 1})\,$, $16\times({\bf 1}, {\bf 1}, {\bf 1}, \frac{1}{2}, -\frac{1}{2},{\bf 1},{\bf 1})\,$\\
				\hline			
				$b_3+T_2+x (+z_1)$ & $4\times({\bf 6}, {\bf 1}, {\bf 1}, \frac{1}{2}, \frac{1}{2},{\bf 1},{\bf 1})\,$, $4\times({\bf 1}, {\bf 1}, {\bf 1}, \frac{1}{2}, \frac{1}{2},{\bf 6},{\bf 1})\,$, $16\times({\bf 1}, {\bf 1}, {\bf 1}, \frac{1}{2}, -\frac{1}{2},{\bf 1},{\bf 1})\,$\\
				\hline
				$b_3+T_1+T_2+x$ & $4\times({\bf 1}, {\bf 2}, {\bf 2}, \frac{1}{2}, \frac{1}{2},{\bf 1},{\bf 1})\,$\\
				\hline
				$b_3+\alpha (+ z_1)$ & $4\times({\bf 4}, {\bf 1}, {\bf 1}, 0, 0,{\bf \bar{4}},{\bf 1})\,$\\
				\hline
				$b_3+T_1+T_2+\alpha  (+z_1)$ & $4\times({\bf 4}, {\bf 1}, {\bf 1}, 0, 0,{\bf \bar{4}},{\bf 1})\,$\\
				\hline
				$b_3+x+z_1  (+z_1+z_2)$ & $4\times({\bf 1}, {\bf 1}, {\bf 1}, \frac{1}{2}, \frac{1}{2},{\bf 1},{\bf 12})\,$\\
				\hline 
				$b_3+T_1+T_2+x+z_1 (+z_1+z_2)$ & $4\times({\bf 1}, {\bf 1}, {\bf 1}, \frac{1}{2}, \frac{1}{2},{\bf 1},{\bf 12})\,$\\
				\hline
		\end{tabular}}
		\caption{\it Spectrum of massless scalar matter and quantum numbers under the gauge bundle for Model 3. Multiplets of the hidden $SO(6)$ and $SO(12)$ gauge groups are obtained from sector pairs of the form  $\xi (+z_1) = \{\xi, \xi+z_1\}$ and $\xi(+z_1+z_2)=\{\xi, \xi+z_1+z_2\}$ respectively. \label{TT9}}
	\end{table}
	\FloatBarrier
Let us remark that fractionally charged exotic states arise in four types: (i) Doublets of $SU(2)_R$ originating from the sectors  $S+b_1+x+\alpha (+z_1)$, $S+b_1+T_3+x+\alpha (+z_1)$; (ii) Doublets of $SU(2)_L$ arising from
$S+b_2+x+\alpha (+z_1)$, $S+b_2+T_3+x+\alpha (+z_1)$;  (iii) Vectorials of 
$SU(4)$ from the sectors $S+b_3 + T_1+ \alpha (+z_1)$; (iv) Anti-vectorials of  $SU(4)$ from $S+b_3 + T_2+ \alpha (+z_1)$. As seen from Tables \ref{TT10} and \ref{TT9}, they all transform as vectorials/antivectorials of $SO(6)\sim SU(4)$ hidden group factor. This opens up the interesting possibility that  all exotic fractional charged states confine, as the hidden  $SU(4)$  gauge  coupling could grow  rapidly and lead to confinement at scales higher than the TeV scale. A similar scenario has been considered in the case of the supersymmetric flipped $SU(5)$ model \cite{Antoniadis:1989zy,Leontaris:1990bw}, but it had not been implemented before in the context of a Pati--Salam model.

\section{Conclusions}
	
In this work we constructed heterotic theories with Pati--Salam gauge group, where space-time $\mathcal N=1$ supersymmetry is spontaneously broken by the stringy Scherk--Schwarz mechanism. Although Bose-Fermi degeneracy is no longer present in the full tower of string states, it is still possible to preserve it in the massless sector of the theory. The resulting non-supersymmetric theories, known as super no-scale models, enjoy exponentially suppressed values for the effective one-loop potential and depending on its shape, the theory may be dynamically driven to low values for the supersymmetry breaking scale $m_{3/2}$.

Our construction is based on the joint application of the free-fermionic and orbifold formalisms, by exploiting their equivalence valid for toroidal order-2 orbifold groups at special factorized points of moduli space. Modulo trivial degeneracies, out of the initial sample of $\sim 10^{10}$ models, a set of selection criteria is introduced that imposes a series of self-consistency and phenomelogical compatibility requirements, including the presence of chiral matter and the super no-scale condition. The latter truncate the model space down to approximately one million configurations which are subsequently classified according to the structure of their one-loop effective potentials. We observed that the salient features of the effective potential may be described in terms of two parameters, termed $\lambda_1$ and $V_F$, which depend on the mismatch in degeneracies between bosons and fermions at the first mass level, as well as on the value of the effective potential at the fermionic point.

The scan of the models we perform is exhaustive and their classification allows for a direct comparison with parent models based on $SO(10)$ gauge symmetry, studied in \cite{Florakis:2016ani}. The latter are obtained from the models studied here by removing a single basis vector from the defining basis of the fermionic construction. The resulting observation is that the super no-scale property, as well as the main features of the effective potential are not preserved by this map. This is not surprising since, typically, the addition of a basis vector (equiv., the introduction of an additional $\mathbb Z_2$ orbifold factor) drastically effects the spectrum of the theory.

The models constructed in this work were based on ascribing the same boundary conditions to pairs of compexified fermions. This choice was made for technical reasons, both for keeping the sample space to a range allowing for an exhaustive computer scan, as well as because the map between the free-fermionic and orbifold descriptions is then significantly simpler. The price to pay for this explicit tractability, is the fact that compefixied fermions introduce additional degeneracies in the chiral matter of the theory, and the number of generations is constrained to appear in multiples of four. It is an interesting and important problem to extend our formalism to the case where the additional $U(1)$ symmetry of the complexification is broken by suitable (real) boundary conditions, such that the resulting theories be compatible with three generations of chiral matter. We hope to return to this issue, as well as the question of the running of gauge couplings and the scale of string unification in future work.

\section*{Acknowledgments}
	
J.R. would like to acknowledge participation in the COST Association Action CA18108 ``Quantum gravity phenomenology
in the multi-messenger approach". The research of K.V. is co-financed by Greece and the European Union (European Social Fund - ESF) through the Operational Programme ``Human Resources Development, Education and Lifelong Learning" in the context of the project ``Strengthening Human Resources Research Potential via Doctorate Research" (MIS-5000432), implemented by the State Scholarships Foundation (IKY).

\bibliographystyle{JHEP1}




\providecommand{\href}[2]{#2}\begingroup\raggedright\endgroup

\end{document}